\documentclass[aps,prc,twocolumn,groupedaddress,nofootinbib]{revtex4-1}

\usepackage{amssymb}
\usepackage{graphicx}
\usepackage{graphics}
\usepackage{booktabs}
\usepackage{siunitx}
\usepackage{chemmacros}
\usepackage{enumitem}
\usepackage{xcolor}
\usepackage{braket}
\usepackage{bbold}
\usepackage{subcaption}
\usepackage{ulem} 

\newcommand{\be}{\begin{equation}} 
\newcommand{\ee}{\end{equation}}    
\newcommand{\bc}{\begin{cases}}     
\newcommand{\ec}{\end{cases}}

\begin{document}

\title{Electric dipole response of $sd$-shell nuclei within the Configuration-Interaction Shell Model approach}
\author{O.~Le Noan}
\author{K.~Sieja}

\address{Universit\'e de Strasbourg, IPHC, 23 rue du Loess 67037 Strasbourg, France\\
CNRS, UMR7178, 67037 Strasbourg, France}
\date{today}
\begin{abstract}
Reliable theoretical predictions of nuclear dipole excitations are crucial for various nuclear applications, particularly in nuclear astrophysics. Calculations of radiative capture cross sections often rely on theoretical $\gamma$ strength functions, with the electric dipole response being the dominant component. Recent experimental and theoretical efforts in this area have focused on nuclei with mass numbers $A\le60$ which appear of interest to model intergalactic propagation of ultra-high-energy cosmic rays.
We aim at a systematic description of the $E1$ strength 
of nuclei with mass numbers between 17 and 40 of interest for various applications. Additionally, we seek to better understand the nature of the low-energy dipole strength in neutron-rich nuclei, known as pygmy dipole resonances.
We use the Configuration Interaction shell-model framework in the $p-sd-pf$ valence space with a previously established empirical Hamiltonian. Full $1\hbar\omega$
configuration space is considered to describe $E1$ excitations. 
Systematic calculations are carried out for the $sd$-shell nuclei
and compared to available data and other models. The nature of different resonances is assessed based on calculations of transition densities and analysis of the wave functions.  
Systematic results of photoabsorption strength show good agreement with experimental data, provided a renormalization of the dipole operator is applied to account for the correlations outside the model space. 
Transition densities are computed in $^{26}$Ne and confirm the pure isovector character of the Giant Dipole Resonance. The strength at 7-10MeV is shown to have a distinct structure from the giant resonance and from the lowest-energy excitations, with largely fragmented wave functions and 
transition densities of isovector character at the edge of the nucleus. 
The Configuration-Interaction shell model is proved to be a valuable tool in the description of the photoresponse of light nuclei, providing more accurate results than the usually employed approaches. It gives additional and valuable insight into the nature of low-energy dipole strength, supporting the interpretation of the pygmy dipole strength as an oscillation of the neutron skin against the proton-neutron core.

\end{abstract}

\maketitle

\section{Introduction}
Photo-nuclear reaction rates serve as pivotal parameters in various applications
of nuclear physics, acting as
fundamental probes of nuclear structure, from single particle dynamics to collective excitations. These reactions
reveal the nature of complicated correlations inherent to nuclear systems. 
Notably, the $E1$ dipole resonance propelled by the external fields stands out for its significance. It is dominated by the Isovector Giant Dipole Resonance (IVGDR) depicted as
a collective oscillation of protons against the neutrons within nuclei. In neutron-rich nuclei, an oscillatory motion of  
the neutron-excess against the proton-neutron core was predicted to give rise to low-lying
$E1$ strength known as the Pygmy Dipole Resonance (PDR) \cite{Mohan71}. The low-energy excess of $E1$ strength was observed in a number of nuclei, see e.g. \cite{Savran2013} 
and references therein, though its theoretical interpretation
is still debated \cite{Paar2005,Vretenar2012,reinhard_information_2013}. The knowledge of the PDR serves to probe the neutron-skin thickness of medium to heavy nuclei \cite{Tamii2011, Piekarewicz2011},
constrain the nuclear symmetry energy \cite{Roca-Maza2015},
and the properties of neutron stars \cite{Horowitz2001}.

Applications reliant on precise knowledge of photo-nuclear reactions 
span a broad spectrum, encompassing astrophysical nucleosynthesis processes like 
the $r$-process and the intergalactic propagation of ultra-high-energy cosmic rays (UHECR). 
While the former requires the knowledge of radiative decay strength functions 
for about 5000 nuclids, involving heavy and very neutron-rich nuclei, 
the latter is particularly concerned with 
nuclei with masses
$A\le60$ not far from the stability. PANDORA (Photo-Absorption of Nuclei and Decay Observation for Reactions in Astrophysics) 
project has recently been proposed to explore the photoresponse of light nuclei comprehensively \cite{PANDORA}. 
It entails a multifaceted approach integrating experimental measurements and theoretical predictions of photon strength functions (PSF). Given its focus on light and mid-mass nuclei, Configuration Interaction Shell Model (CI-SM) emerge as a fitting theoretical tool to delineate the photoresponse and furnish necessary predictions.

The calculations of electric dipole excitation strength functions within the CI-SM
are not abundant in the literature due to the large sizes of Hamiltonian matrices involved 
in such calculations and difficulties related to derivation of multi-shell valence space interactions, see e.g. \cite{Sagawa99, ma_proton_2012, Utsuno2015, Togashi2018, Sieja-PRL} for available examples. Recently, we have studied the photoabsorption and decay strength functions
of Ne isotopes within the CI-SM framework in order to probe the viability of the Brink-Axel hypothesis in the region of PDR \cite{Sieja-PRL2}. 
Building upon these endeavours, this study aims to broaden the scope by systematically analyzing the $E1$ response of $sd$-shell nuclei, comparing to experimentally known cases and other available theoretical calculations. 
Furthermore, we focus on the $E1$ sum rules and distributions and their dependence on the correlations included in the model. We determine the effective charge to be applied on the isovector dipole operator mapped into the considered valence space
to enhance the theoretical description of $E1$ strength in $sd$-shell nuclei within the CI-SM framework. Finally, we revisit the dipole response of Ne isotopes. We study in detail the structure of the states in the GDR and PDR regions and examine the proton and neutron transition densities in $^{26}$Ne. While proton-pygmy resonances in $^{17,18}$Ne were previously discussed in the shell-model framework in Ref. \cite{ma_proton_2012}, our emphasis here is on the neutron PDR.

Our paper is organized as follows: in Section \ref{CI-SM} we discuss our CI-SM calculations, model space and interaction used and our computational methods.
We also remind basic definitions used to characterize dipole strength functions. In Section \ref{Hamil} we discuss the renormalization of the dipole operator. 
The comparison of the experimentally known cases to our CI-SM calculations
is summarized in Sec. \ref{SYS} and discussed in the context of available theoretical calculations. 
In Section \ref{Neon} we present a detailed analysis of the $E1$ dipole response in the Ne chain and focus on the 
PDR in $^{26}$Ne. The conclusions of this work are collected in Sec. \ref{CONC}.
 
\section{Theoretical framework \label{CI-SM}}
\begin{figure}
\includegraphics[width=0.5\textwidth]{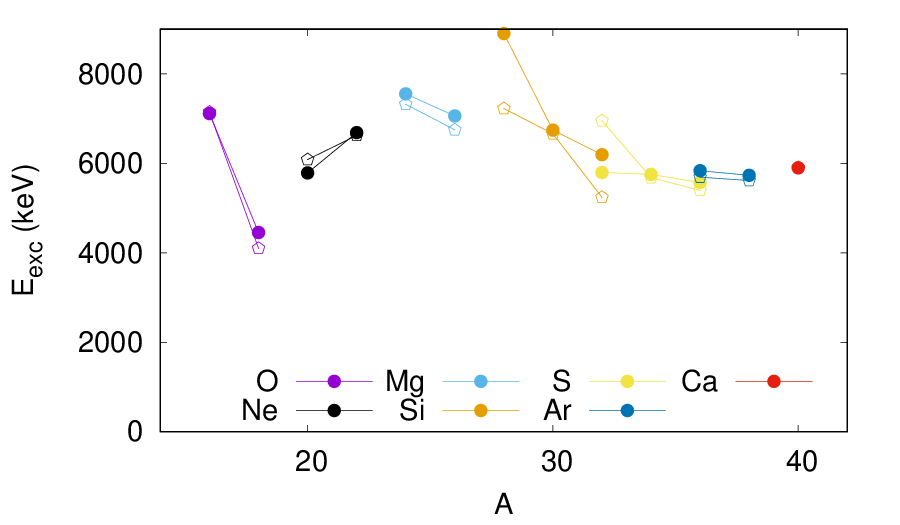}
\caption{Excitation energies of the lowest $1^-$ states in even-even $sd$-shell nuclei. Experimental results from 
\cite{NNDC} (filled symbols) are compared to shell-model calculations with the PSDPF interaction (open symbols).\label{fig-spec}}
\end{figure}

In this section we discuss the details of our theoretical framework and we remind the definitions of all the quantities that are discussed in the article.  
The CI-SM, known as well as large-scale shell-model approach, permits a digonalization of the (generally) one- plus two-body nuclear Hamiltonian within the configuration space that can be formed by placing nucleons within a given set of single-particle orbits, called the model space. In our case a full $1\hbar\omega$ $p-sd-pf$ model space is used to study the spectroscopy and $E1$ transitions of the $sd$-shell nuclei. Selected results are also verified extending the calculations to $(1+3)\hbar\omega$ excitations for negative parity and $(0+2)\hbar\omega$ for the positive parity states.
The shell-model Hamiltonian reads:
\begin{equation}
H=\sum_i \epsilon_i c_i^\dagger c_i+ \sum_{ijkl}V_{ijkl}c_i^\dagger c_j^\dagger c_l c_k+\beta H_{COM}
\end{equation}
where the center-of-mass (COM) Hamiltonian with a multiplication coefficient $\beta=10$ is added to push up the
COM eigenvalues to the energy range not considered here \cite{gloeckner_spurious_1974}.
The isovector $E1$ transition operator is considered
\begin{equation}
\hat O_{1\mu}=-e\frac{Z}{A}\sum_{i=1}^N r_i Y_{1\mu}(\hat r_i)+e\frac{N}{A}\sum_{i=1}^Z r_iY_{1\mu}(\hat r_i)
\end{equation}
substracting the COM motion. 
The effective interaction used in this work is the semi-empirical one developed in \cite{Mouna2011}
to describe intruder states in the $sd$-shell and employed later for a systematic study 
of $E3$ transitions in $sd$-shell nuclei \cite{Mouna2017}. It will be dubbed hereafter as PSDPF.  This interaction reproduced the known negative-parity levels in the $sd$-shell nuclei with the rms deviation value of 400keV. 
In Fig. \ref{fig-spec} we show how the PSDPF interaction performs for the lowest $1^-$ states in the even-even $sd$-shell nuclei. Those are calculated as one particle - one hole (1p-1h) excitations from the $p$ to $sd$ and from the $sd$ to the $pf$ shells. The agreement is satisfying, though much better at the beginning and at the end of the shell than in its middle. The rms deviation for 15 states shown in the Figure is 600keV. One should however note the worse deviation from the experiment concerns $^{28}$Si and $^{32}$Si and in both nuclei the second excited $1^-$ shell-model state fits much better the experimentally known value. 

We emphasize the fact that the PSDPF interaction was adjusted to reproduce low-lying excited states while we use it here to compute systematically the $E1$ response which involves much higher excitations. Distributions of $B(E1)$ strengths shown in Secs. \ref{SYS} and \ref{Neon} are computed using the Lanczos strength functions method
which permits getting the strength per energy interval in an efficient way \cite{RMP}. 
We remind that the choice of the starting vector, called pivot, used in the Lanczos diagonalization procedure is arbitrary.
Given a transition operator $\hat O$ one can define a pivot of the form $\hat O |\Psi_i\rangle$,
where $|\Psi_i\rangle$ can be chosen as any shell-model state, and carry on Lanczos diagonalization.
The unitary matrix $U_{ij}$ that diagonalizes the Hamiltonian after N Lanczos iterations contains then   
in its first row the amplitude of the pivot in the $j$th eigenstate. Thus $U_{1j}^2$ as a function of 
eigenergies $E_j$ defines the strength function of the pivot state.
Note that to obtain the total strength $S_0$ for the ground state 
only diagonalization of one state has to be carried out, 
as the sum rule is the norm of the pivot state obtained 
by acting with the transition operator on the initial state.
The remaining moments are extracted from the discrete 
distributions obtained with the Lanczos strength function method with 300 iterations.
These calculations are done using the $m$-scheme shell-model code ANTOINE \cite{ANTOINE, RMP}.

To characterize the $E1$ strength distributions we analyze sum rules, centroids and widths 
obtained following standard definitions (see e.g. \cite{Ring80, Stetcu2003}). We consider the total strength:
\begin{equation}
S_0=\sum_\nu|\langle\nu|\hat O|0\rangle|^2,
\label{NWSR}
\end{equation}
being the non energy-weighted sum rule (NWSR) of $\hat O$ transition operator, where $\ket{0}$ is the nuclear ground state and $\ket{\nu}$ an excited eigenstate. 
The centroid and width are defined as
\begin{eqnarray}
\bar S=\frac{S_1}{S_0},\quad \Delta S=\sqrt{\frac{S_2}{S_0}-\bar S^2},
\end{eqnarray}
where:
\begin{equation}
S_k=\sum_\nu(E_\nu-E_0)^k |\langle\nu|\hat O|0\rangle|^2 
\end{equation}
is the sum rule of the order k, with $E_\nu-E_0\equiv E_\textrm{exc}(\nu)$ the excitation energy of the $\ket{\nu}$ state. The $S_1$ sum rule is referred to hereafter as Energy Weighted Sum Rule (EWSR). 

The reduced transition probability is calculated as 
\begin{equation}
B_{\nu 0}=\frac{1}{2J_0+1}\langle \nu||\hat O|| 0\rangle^2\,, 
\end{equation}
The $B(E1)$ distributions are additionally convoluted with Lorentzians of width 
$\Gamma/2$=0.5~MeV:
\begin{equation}
S(E)=\sum_\nu B_{\nu 0}\frac{1}{\pi}\frac{\Gamma/2}{(E-E_\textrm{exc}(\nu))^2+(\Gamma/2)^2},
\label{Eq-SE1}
\end{equation}
which leads to continuous strengths presented in selected figures.
We also present our results as PSF (in ~MeV$^{-3}$ units)
defined by the relation \cite{Bartholomew}:
\begin{equation}
f_{E1}(E)=\frac{16\pi}{27(\hbar c)^3}S(E)\,.
\end{equation}
When dealing with nuclear dipole response a convenient approximation of the EWSR is the Thomas-Reich-Kuhn (TRK) sum rule, given by \cite{mottelson_nuclear_1969}:
\begin{equation}
    S_1^{\scriptscriptstyle TRK} = \frac{9\hbar^2 e^2}{8\pi m} \frac{NZ}{A} = 14.8\frac{NZ}{A} \quad e^2\textrm{fm}^2~\textrm{MeV.}
    \label{TRK}
\end{equation}

Following \cite{traini_study_1987} we can write the EWSR in terms of the enhancement factor $K$, which accounts for the non-local contribution to the sum rule:
\begin{equation}
    S_1 = S_1^{\scriptscriptstyle TRK}(1+K).
    \label{Eq-enh}
\end{equation}
It is known \cite{lipparini_sum_1989} that the experimental EWSRs critically depend on an energy cutoff $E_\gamma^{\textrm{max}}$. 
Typically for $E_\gamma^{\textrm{max}} \sim 140$ MeV, $\mathcal{O}  (K) = 1$ while if $E_\gamma^{\textrm{max}} \sim E_{\scriptscriptstyle GDR}$, $K$ is much smaller. In the present calculations, we computed the EWSR typically in the $E_\gamma^{\textrm{max}}\le 50$ MeV region: though the Lanczos strength function procedure allows us us to cover the whole energy range spanned by 1p-1h excitations, there are hardly any non-zero $B(E1)$s connecting to higher excited states.   


Further we recall the definition of dipole polarizability, a quantity considered of interest to extract the neutron-skin thickness \cite{Roca-Maza2015}. Dipole polarizability is related directly to the inverse-energy-weighted sum rule $S_{-1}$\cite{lipparini_sum_1989, Ring80}:  
\begin{equation}
\alpha_D=\frac{8\pi}{9}S_{-1}=\frac{\hbar c}{2\pi^2}\sigma_{-2},
\end{equation}
where $\sigma_{-2}$ stands for the (-2) moment of the total photoabsorption cross
section $\sigma_{tot} (\varepsilon_\gamma)$ defined as:
\begin{equation}
\sigma_{-2}=\int_{0}^{E_\gamma^{\textrm{max}}} \frac{\sigma_{tot}(E_\gamma)}{E_\gamma^2}d E_\gamma.
\end{equation}
$E_\gamma^{\textrm{max}}$ is taken 50~MeV in theoretical calculations. The results in this work are expressed in terms of $\sigma_{-2}$ rather than $\alpha_D$, to be directly comparable to similar shell-model calculations presented recently in Ref. \cite{Orce2023}.

In the analysis of pygmy resonances in Sec. \ref{neon26} we use transition densities between final and initial states defined as
\begin{equation}
\delta \rho_{\nu 0}(\vec r)=\langle \nu|\sum_k\delta(\vec r-\vec r_k)|0\rangle\,,    
\end{equation}
(see Ref. \cite{ma_proton_2012} for details of calculations in the shell-model context). 
To obtain a quantitative measure of dipole collectivity of the calculated states we examine two additional indicators: the cumulative sum \cite{lanza_multiphonon_2009} and the configuration-space entropy. We can rewrite the reduced transition probability as 
\begin{align}
    B_{\nu 0} = \bigg | \sum_{k_\alpha k_\beta} \big ( X^{p}_{k_\alpha k_\beta} + X^{n}_{k_\alpha k_\beta} \big ) \bigg |^2,
    \label{eq-cumul}
\end{align}
where 
\begin{align}
    X^{\tau}_{k_\alpha k_\beta} = \frac{e_{\textrm{eff}}^\tau D_{k_\alpha k_\beta}^\tau}{\sqrt{2J_0+1}}\bra{l_\alpha j_\alpha}|Y_1 |\ket{l_\beta j_\beta}\bra{n_\alpha l_\alpha}R\ket{n_\beta l_\beta}
\end{align}
and $k_\alpha = (n_\alpha l_\alpha j_\alpha)$ define the single-particle quantum numbers in the harmonic oscillator basis. $e_{\textrm{eff}}^p=eN/A$, $e_{\textrm{eff}}^n=-eZ/A$ and $D_{k_\alpha k_\beta}^\tau$ stands for the one-body transition density (OBTD)\cite{ma_proton_2012}. Then the cumulative sum is defined as $\sum_{k_\alpha k_\beta} \big ( X^{p}_{k_\alpha k_\beta} + X^{n}_{k_\alpha k_\beta} \big )$. In a fully collective state all single-particle contributions $X^{\tau}_{k_\alpha k_\beta}$ add up constructively, leading to an enhanced $B(E1)$ value. 

The CI-SM eigenstates are expanded in the basis of Slater determinants (configurations) $\{ \ket{n}\}$: 
\begin{eqnarray}
\ket{\psi} = \sum_n c_n^\psi \ket{n}\,.
\end{eqnarray}
To characterize the spread of the wave-functions among various configurations we introduce the configuration entropy in the $|\psi \rangle$ state:\footnote{The analogy can be made to the usual Von Neumann entropy $S(\rho)=-\textrm{Tr}(\rho \ln \rho)$ if one considers Slater determinants $\ket{n}$ as playing the role of eigenstates $\ket{\psi}$ in the general density matrix $\rho= \sum_\psi \eta_\psi \ket{\psi}\bra{\psi}$.} 
\begin{align}
    S_\psi = - \sum_n |c_n^\psi|^2\ln{|c_n^\psi|^2}
\end{align}
such that $S_\psi = 0$ for a state formed by a single determinant and the maximum entropy $S_{\textrm{max}}= \ln{d} $ with $d$ being the $m$-scheme dimension. 

\section{Renormalization of the electric dipole operator\label{Hamil}}
Systematic calculations of the $E1$ strength on $sd$-shell nuclei carried in this work (see next Section)
reveal rather constant deviations from the available experimental data,  
overestimating systematically the computed photoabsorption cross sections which appear common to previous calculations of the same type: 
for carbon \cite{Suzuki_E1} and oxygen isotopes \cite{Sagawa99} with the WBP family of interactions \cite{WBP}, for $p$-shell and $sd$-shell nuclei calculated with WBP and FSU interactions in Ref. \cite{Orce2023}, the $N=Z$ nuclids computed in the $p-s-d_{5/2}$ model space in Ref. \cite{Lu2018} (even though the model space is not full $1\hbar\omega$ in this case). 
Overestimation of the photoabsorption cross section is also present in $1\hbar\omega$
shell-model and Monte-Carlo Shell Model calculations of heavier nuclei: $^{48}$Ca
\cite{Utsuno2015}, $^{90-94}$Zr and $^{124}$Sn \cite{Togashi2018}.
It appears thus a generic feature of the $1\hbar\omega$ shell-model calculations of the $E1$ strength, independent of the details of calculations. Among the plausible explanations proposed in the literature for this behaviour of CI-SM results one finds:
\begin{enumerate}[label=(\roman*)]
   \item too strong particle-hole content of the effective interaction \cite{Orce2023},
    \item an insufficiently large model space due to $1\hbar \omega$ truncation; the $(1+3)\hbar\omega$ calculations were reported to bring 15-30\% reduction of the sum rule
\cite{Sagawa99, Utsuno2015, Shimizu_Comex, Sieja-PRL2}. 
\end{enumerate}

\begin{figure}
\begin{center}
\includegraphics[width=0.5\textwidth]{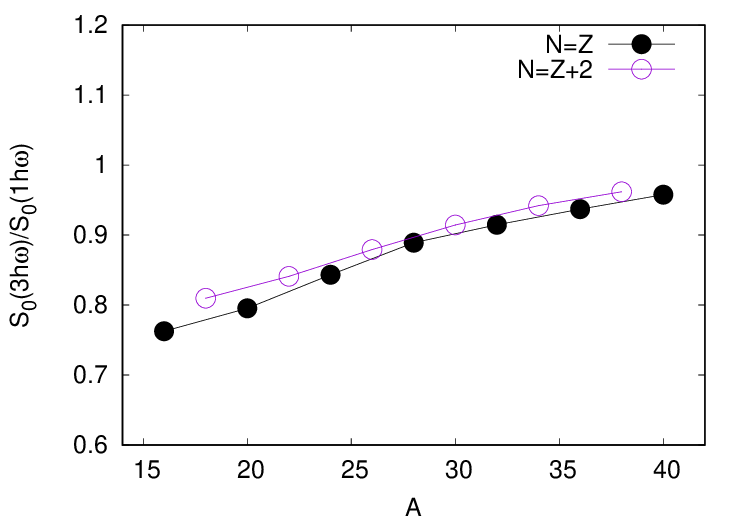}
\end{center}
\caption{Ratio of the total $E1$ strength $S_0$ computed with the PSDPF interaction in $3\hbar\omega$ and $1\hbar\omega$ model spaces as function of nuclear mass $A$.
The calculations are performed for even-even $N=Z$ and $N=Z+2$ $sd$-shell nuclei.\label{fig-sr}}
\end{figure}

To illustrate the latter point, we show the variation of the total strength $S_0$ as a function of the chosen configuration space. The calculations for a set of $sd$-shell even-even nuclei were performed using different p-h excitations. 
Initially, we considered the ground state within the 0$\hbar\omega$ configuration and for $1^-$ states we allowed for all possible 1p-1h excitations, referred to as $1\hbar\omega$ calculations. Further, we included  excitations within (0+2)$\hbar\omega$ space for the $0^+$ ground state and (1+3)$\hbar\omega$ for the excited states, leading to the $3\hbar\omega$ configuration space. The effect of extra excitations is well visible in Fig. \ref{fig-sr}, where the ratio of the sum rules computed in various configuration spaces as a function of mass number is presented. 
We note a reduction of the total strength between 22\% and 1\%, decreasing with the mass, and thus the number of hole states in the $sd$ shell. The observed reduction in the sum rule is attributed to enhanced configuration mixing in the ground states, in which only $60-65\%$ of the wave functions resides in the $0\hbar\omega$ model space in enlarged calculations. The $2\hbar\omega$ excitations cause greater interference of the dipole matrix elements between the ground and excited states. Notably, the figure reveals a clear trend where the reduction effects are most pronounced in nuclei at the beginning of the $sd$-shell. As the mass number increases, $p-sd$ particle-hole excitations become increasingly Pauli-blocked, reducing the $3\hbar\omega$ component in the sum rule states. 

\begin{figure}
\begin{center}
\includegraphics[width=0.5\textwidth]{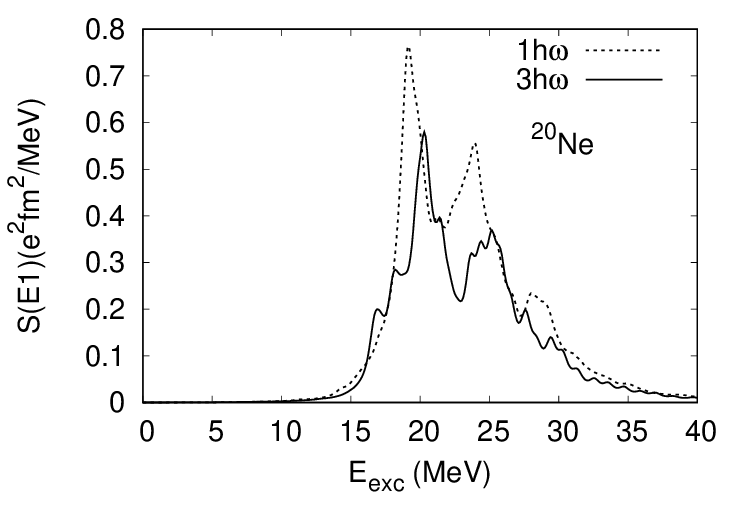}
\end{center}
\caption{Microscopic strength distribution (Eq.\ref{Eq-SE1}) computed in 
$1\hbar\omega$ and $(1+3)\hbar\omega$ model spaces in $^{20}$Ne. \label{fig-ne20}}
\end{figure}

In Fig. \ref{fig-ne20} we depict the $B(E1)$ distributions computed in $1\hbar\omega$ and $3\hbar\omega$ model spaces in $^{20}$Ne : Since the interaction parameters were optimized at the 1p-1h excitation level, we applied a downward shift of 1.6 ~MeV to the $3\hbar\omega$ distribution. This adjustment was made to align the centroids of both calculations in the energy range shown. The primary impact of including $3\hbar\omega$ excitations is then a reduction in the total strength of the $B(E1)$ distribution. Despite this reduction in strength, the overall features and shape of the distribution remain largely unchanged.
As we will show in the next Section, the centroids and shapes of the $E1$ strengths are reproduced to a good accuracy in $1\hbar\omega$ calculations. Additionally, the low-energy levels of opposite parity were included in the PSDPF fit on the $1\hbar\omega$ level only, thus performing systematically $3\hbar\omega$ calculations without refitting of the interaction is not the best option to improve the theoretical description of the $E1$ strengths (the increasing dimensions of the matrices to be diagonalized being an additional drawback). 
The results shown in Fig. \ref{fig-ne20} justify the use of an effective charge, as usually done in the case of other transition operators within CI-SM \cite{RMP}, to account for the core-polarization and other low-order diagrams appearing in the perturbation expansion of the effective operator. Many authors carried out calculations of effective operators since the earliest attempts of using realistic potentials in CI-SM calculations, though the major efforts for electromagnetic transitions went to $M1$ and $E2$ modes, see \cite{Coraggio2020} and references therein.    
In this Section we will thus extract appropriate effective charges for the CI-SM calculations from comparison of our results to available data in light nuclei.
We note that $E3$ transitions were examined in \cite{Mouna2017} with the present PSDPF interaction, leading to a recommended set of effective charges for the $E3$ operator to be used in $1\hbar\omega$ calculations. However, the lack of sufficient experimental data on low-energy $E1$ transitions in $sd$-shell nuclei prevents the application of a similar optimization procedure. The most comprehensive dataset that can be described by the current model is for $^{26}$Mg \cite{schwengner_dipole_2009}, which includes 5 identified $1^-$ states below 10.1 ~MeV, with a total $B(E1;1^-\rightarrow0^+)$ strength of $79.9(66) \times 10^{-4}$ e$^2$fm$^2$. In comparison, CI-SM predicts 9 states within this energy range, yielding a total $B(E1)$ strength of $200 \times 10^{-4}$ e$^2$fm$^2$, suggesting that the charges would need to be reduced by a factor of approximately 0.65 to match the experimental data. 
\begin{figure}
\begin{center}
\includegraphics[width=0.45\textwidth]{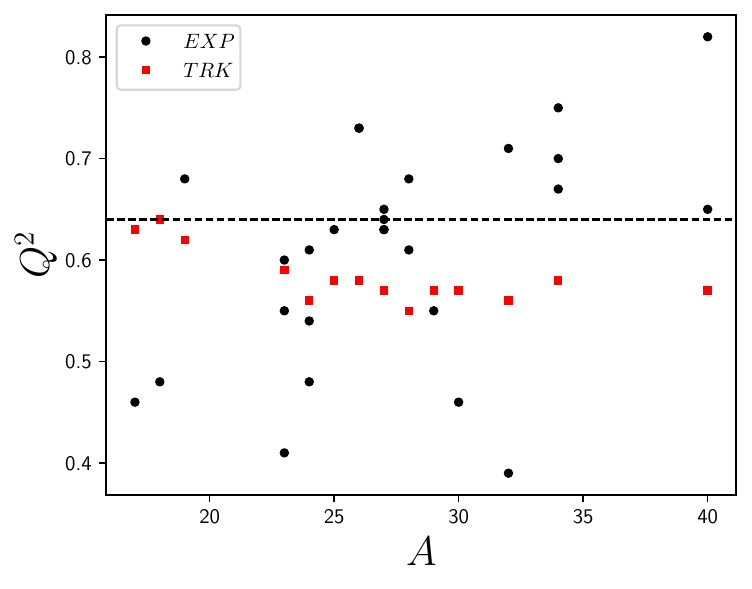}
\end{center}
\caption{Reduction factors obtained as ratio of EWSR from photoabsorption data (black dots) or TRK value (red squares) to theoretical EWSR. The dashed horizontal line indicates the adopted value of $0.64$.}
\label{quenching_exp_trk}
\end{figure}
Since our investigation of the $E1$ response extends to higher excitation energies, we compared our calculations with photoabsorption data from the IAEA PSF database \cite{goriely_reference_2019} to constrain the effective charge for the dipole operator.
As we overshoot the data, we introduce a reduction factor to be applied on the electric charge for protons and neutrons, dubbed hereafter $Q$, and we consider its squared value $Q^2$ deduced from comparison to photoabsorption data. To do so, a folding procedure was applied to the theoretical strength distribution within the same energy range and sampling as each dataset. For nuclei with multiple data sets, a separate $Q^2$ factor was derived for each.
The value $Q^2_{\textrm{\tiny EXP}}$ was defined as the ratio of the experimental to theoretical EWSR, i.e. $Q^2_{\textrm{\tiny EXP}}= \frac{S_1^{\textrm{\tiny EXP}}}{S_1^{\textrm{\tiny SM}}}$ . Alternatively, one could use the NWSR ratio, though it yields no significant difference. For comparison, we introduce a reduction factor based on the TRK ratio, $Q^2_{\textrm{\tiny TRK}}$=$\frac{{S_1}^{\textrm{\tiny TRK}}}{S_1^{\textrm{\tiny SM}}}$. Notably, from Eq. \ref{Eq-enh}, $Q^2$ should always exceed $Q^2_{\textrm{\tiny TRK}}$.

Figure \ref{quenching_exp_trk} presents the estimated values. For photoabsorption data, the factor ranges from 0.4 to 0.8 across the $sd$-shell, showing no clear trend. The ratio of theoretical EWSR to TRK is more consistent across the shell. In some cases, $Q^2_{\textrm{\tiny EXP}}$ falls below the $Q^2_{\textrm{\tiny TRK}}$ threshold, possibly due to significant deviations in CI-SM predictions over the dataset’s energy range or inaccuracies in the normalization of the photoabsorption data.
Despite these discrepancies, CI-SM predictions generally succeed in reproducing both the centroids and fragmentation in many cases (see Section \ref{SYS}) while there is often disagreement between various data sets (see Fig. \ref{PSF_SM_EXP_QRPA}), with $^{27}$Al being the only exception. Adjusting $Q^2$ to the four available datasets for $^{27}$Al results in a consistent factor of approximately 0.64, represented by the dashed line in Fig. \ref{quenching_exp_trk}. This value closely aligns with the mean factor from all experimental data (0.61) and from the TRK ratio (0.58).
 With a $Q^2$ factor of 0.64, the predicted enhancement factor $K$ for $^{27}$Al is 0.14, in agreement with Refs. \cite{Harakeh2001, ishkhanov_giant_2021}, which suggests a relatively small $K$ for nuclei with mass $A < 50$. Therefore, unless otherwise stated, CI-SM results in the following sections will use a factor of $Q^2 = 0.64$ for all nuclei, meaning the effective charges for the dipole transitions should be $0.8~N/A$ for protons and $-0.8~Z/A$ for neutrons.
 Nevertheless, our results emphasize the need for further measurements of low-energy $E1$ transitions and PSF in light nuclei. Such data could lead to more precise information on the $E1$ effective charges and their potential isospin dependence, as currently available data is primarily focused on $N\sim Z$ nuclei.    

\section{Dipole response of sd-shell nuclei\label{SYS}}

\begin{figure*}[htbp]
    \begin{center}
    \begin{subfigure}[b]{0.32\textwidth}
        \includegraphics[width=\textwidth]{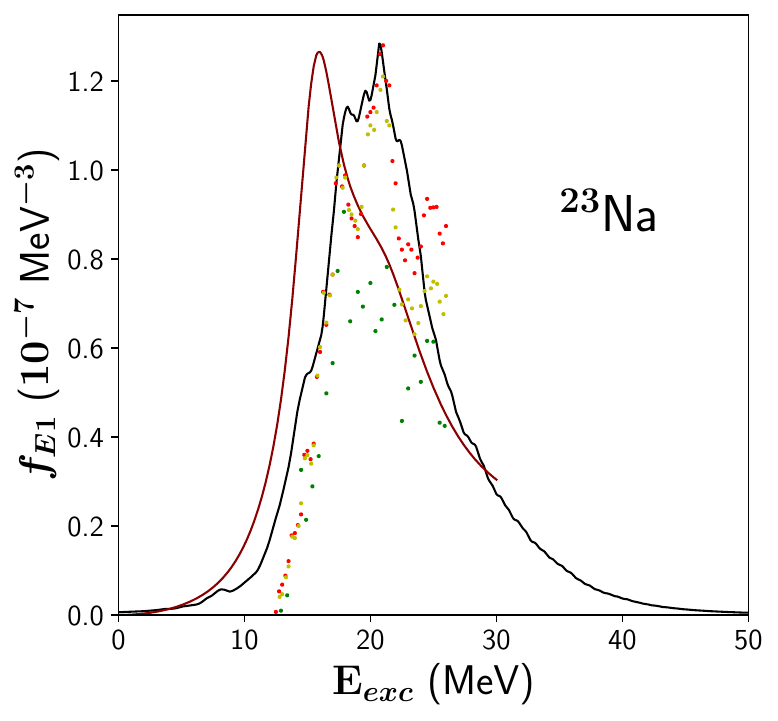}
        \caption{}
        \label{23Na}
    \end{subfigure}
    \hfill
    \begin{subfigure}[b]{0.32\textwidth}
        \includegraphics[width=\textwidth]{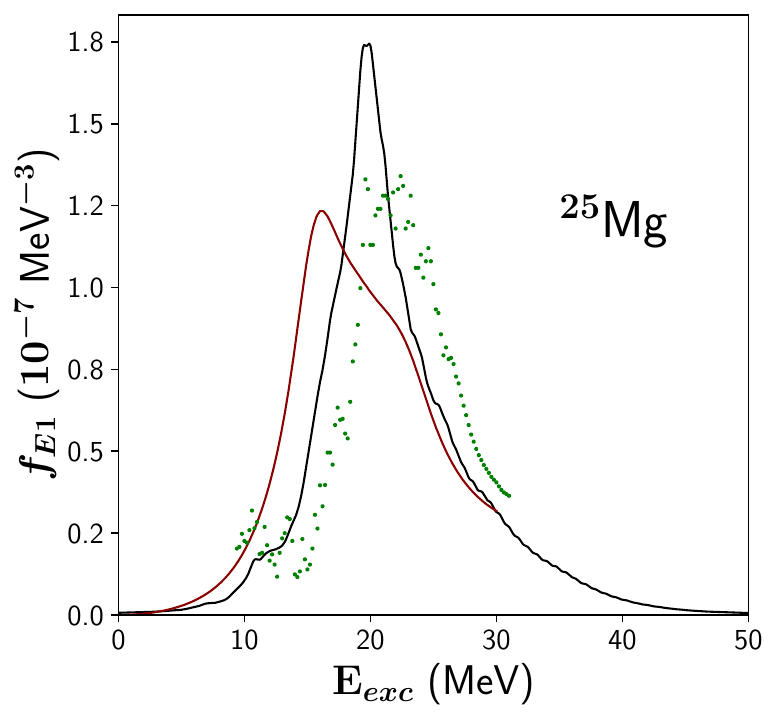}
        \caption{}
        \label{25Mg}
    \end{subfigure}
        \hfill
    \begin{subfigure}[b]{0.32\textwidth}
        \includegraphics[width=\textwidth]{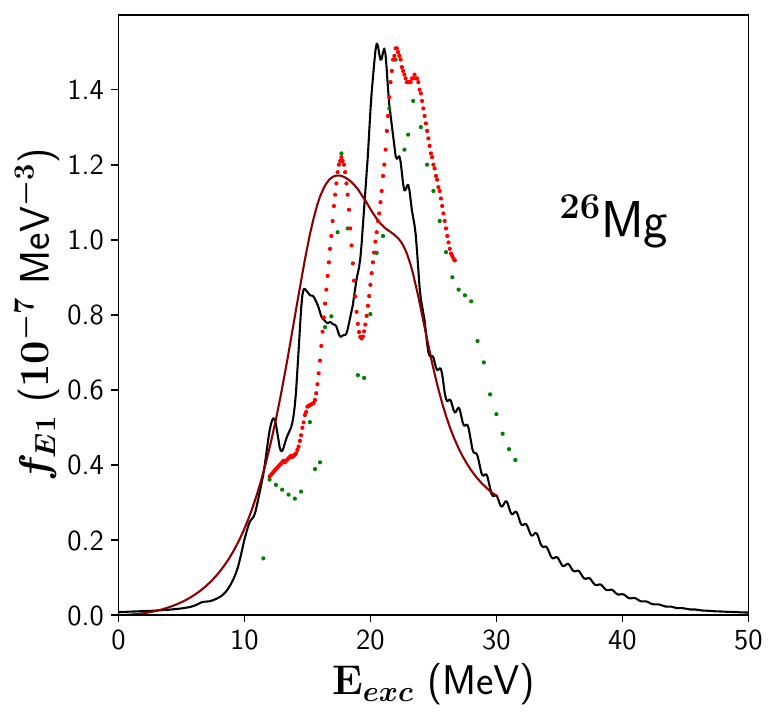}
        \caption{}
        \label{26Mg}
    \end{subfigure}
\\
   \centering
    \begin{subfigure}[b]{0.32\textwidth}
        \includegraphics[width=\textwidth]{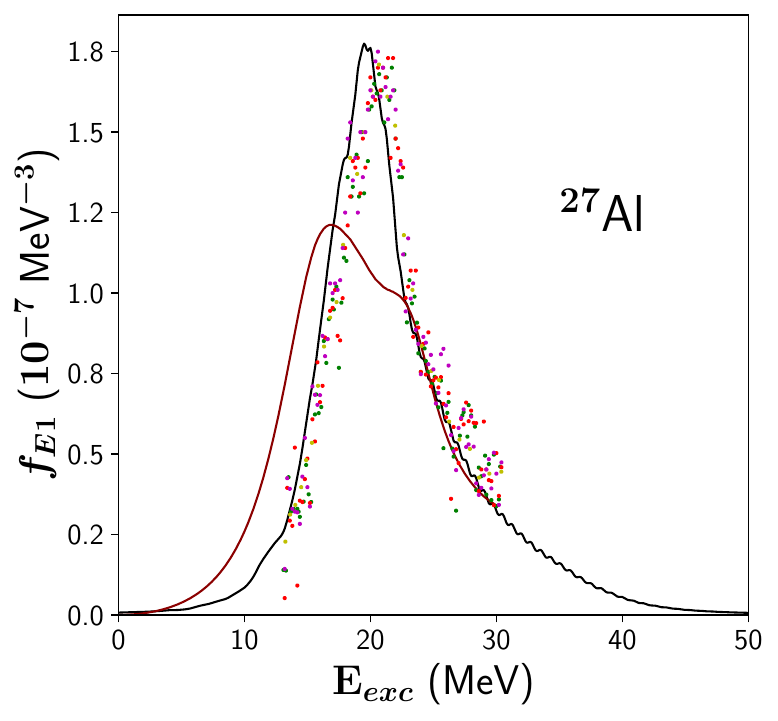}
        \caption{}
        \label{27Al}
    \end{subfigure}
    \hfill
    \begin{subfigure}[b]{0.32\textwidth}
        \includegraphics[width=\textwidth]{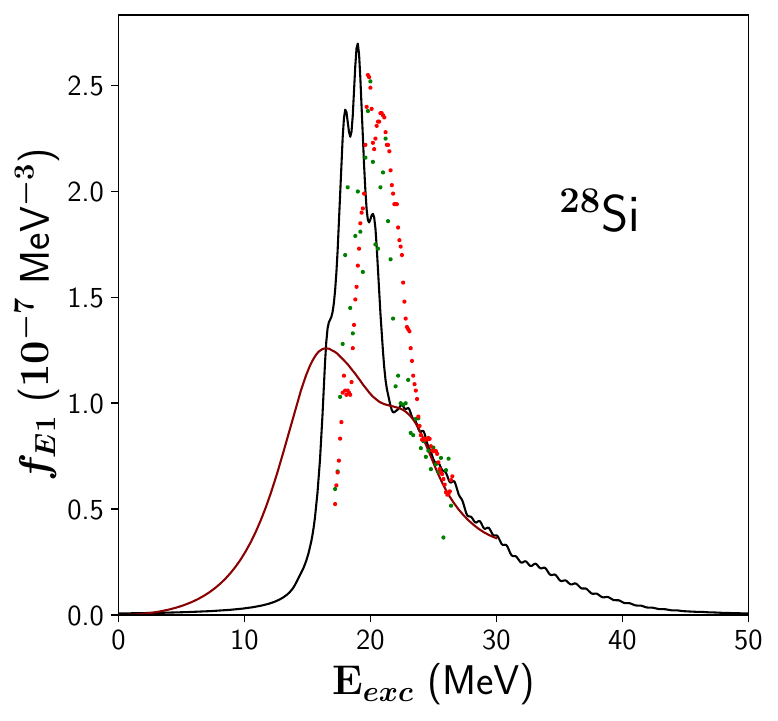}
        \caption{}
        \label{28Si}
    \end{subfigure}
        \hfill
    \begin{subfigure}[b]{0.32\textwidth}
        \includegraphics[width=\textwidth]{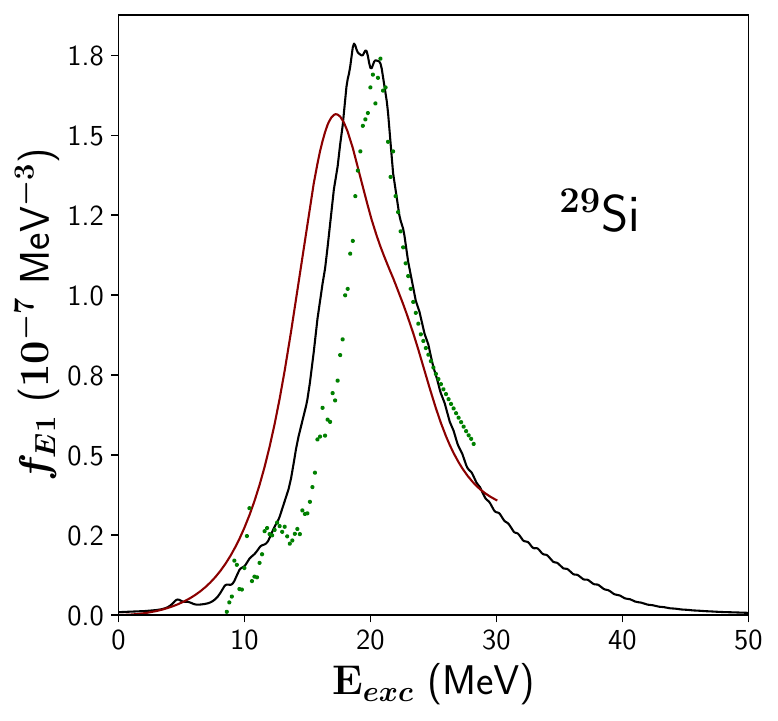}
        \caption{}
        \label{29Si}
    \end{subfigure}
\\
   \centering
    \begin{subfigure}[b]{0.32\textwidth}
        \includegraphics[width=\textwidth]{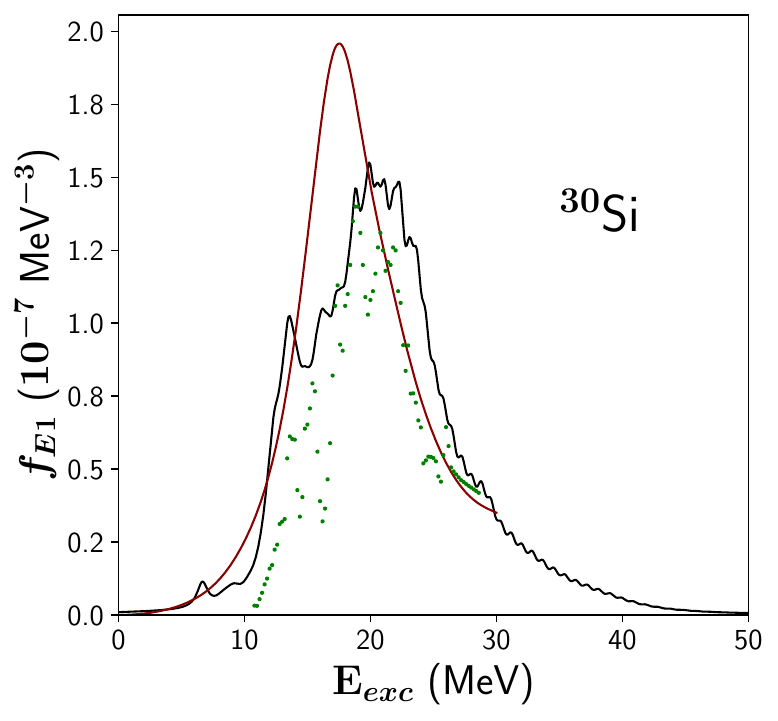}
        \caption{}
        \label{30Si}
    \end{subfigure}
    \hfill
    \begin{subfigure}[b]{0.32\textwidth}
        \includegraphics[width=\textwidth]{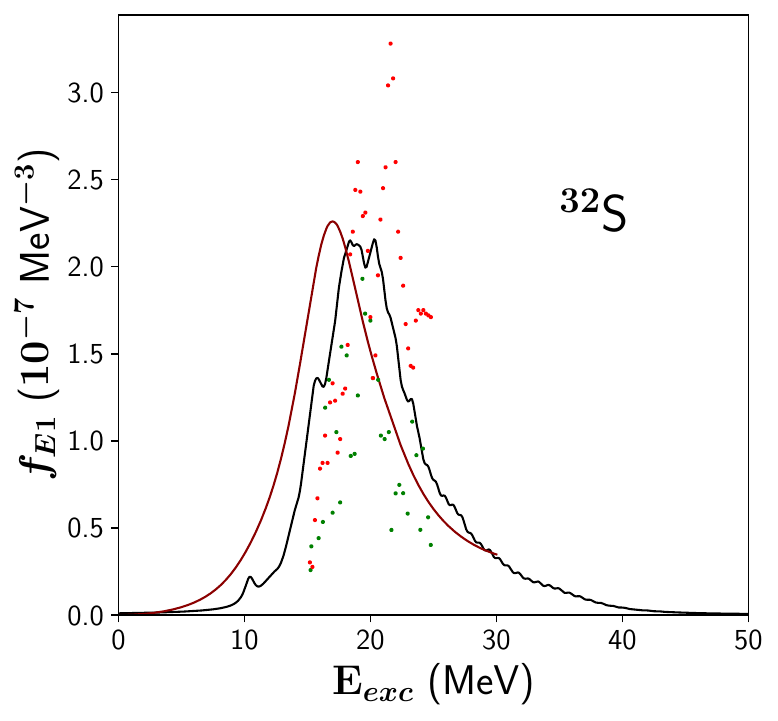}
        \caption{}
        \label{32S}
    \end{subfigure}
        \hfill
    \begin{subfigure}[b]{0.32\textwidth}
        \includegraphics[width=\textwidth]{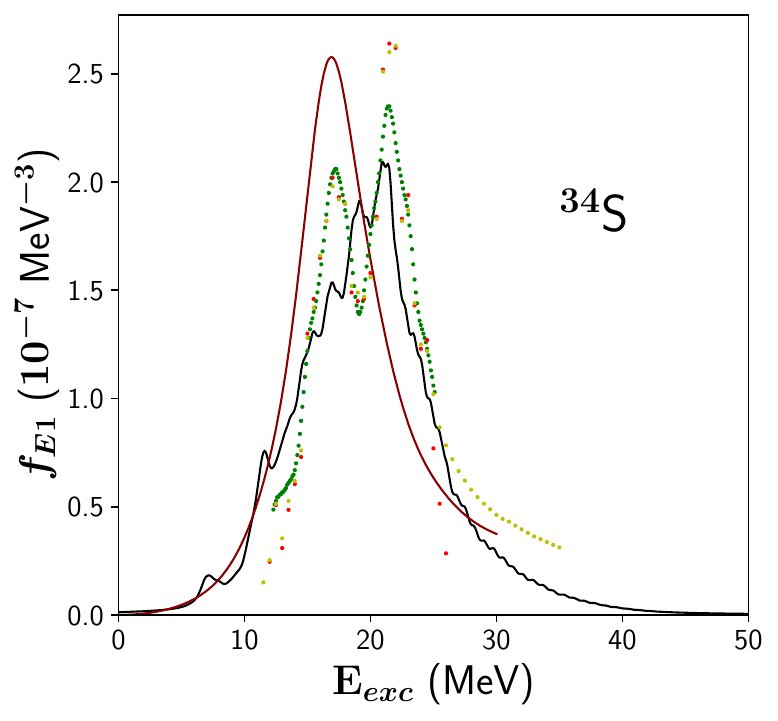}
        \caption{}
        \label{34S}
    \end{subfigure}
    \hfill
        
\caption{Photoabsorption strength functions obtained in this work (black lines) versus available experimental data (points) and 
QRPA results from \cite{Goriely2018} (red lines). Experimental data are taken from the IAEA PSF database \cite{goriely_reference_2019}.}    
\label{PSF_SM_EXP_QRPA}
\end{center}
\end{figure*}
We discuss in this Section systematic photoabsorption strength distributions computed for the $sd$-shell nuclei in the $1\hbar\omega$ model space.
First we show in Fig. \ref{PSF_SM_EXP_QRPA} the obtained PSF for selected cases where experimental data \cite{goriely_reference_2019} is available and compare them additionally to quasiparticle random-phase approximation (QRPA) calculations from \cite{Goriely2018}, which incorporate additional empirical corrections to account for correlations beyond 1p-1h.  
The centroid position and distribution width are well reproduced within the CI-SM framework: the root-mean-square (rms) deviation for 25 centroids is 0.84~MeV, and 0.56~MeV for the width, both calculated within the same energy range as the available experimental data. When excluding nuclei at the extremes of the $p-sd-pf$ valence space, specifically O, F, and Ca, the rms values improve further, reducing to 0.72~MeV for the centroid and to 0.17~MeV for the width. This last number suggests that CI-SM effectively captures the essential correlations needed to model the strength distribution within the experimentally observed region. The shift of the centroid is more cumbersome and would require revision of the effective interaction. However, the CI-SM still outperforms the QRPA model: The QRPA underestimates systematically the position of the centroid, leading to a much larger discrepancy (rms=1.3~MeV), and misses additionally the details of $E1$ distributions, in spite of a similar average error on the widths (rms=0.5~MeV). This is not surprising as the truncation of many-body space of QRPA omits physical effects that are fully accounted for in a complete CI-SM diagonalization and can not be captured by simple empirical shifts. More advanced many-body approaches aim to decrease such errors by including higher order excitations (2p-2h, 3p-3h, phonon-coupling), which enhances the fragmentation of the 
spectrum while shifting the centroid of the   resonance~\cite{Gambacurta2012,Gambacurta2015,Knapp_2023,trippel2016,Tselyaev2020}. 

\begin{table*}
\begin{tabular}{SSSSSSSS}
\hline
\addlinespace[1ex]
{Nucleus} & {$S_0$} & {$S_1^{\scriptscriptstyle TRK}$} & {$S_1$} & {$K$} &{$\bar S $} &{$\Delta S$}  &  {$\sigma_{-2}$ } \\
\addlinespace[1ex]
& {$e^2\textrm{fm}^2$} & {$e^2\textrm{fm}^2 $~MeV} & {$e^2\textrm{fm}^2 $~MeV} & &{~MeV}& {~MeV}& {$\mu b/$~MeV}  \\
\addlinespace[1ex]
\hline
\addlinespace[1ex]
\ch{{}^{17}O} & {2.9} & {62.7} & {64.9} & {0.03} & {22.1} & {4.6} & {568.6} \\
\ch{{}^{18}O} & {3.1} & {65.8} & {66.2} & {0.01} & {21.2} & {6.2} & {660.0} \\
\ch{{}^{18}F} & {3.2} & {66.6} & {71.3} & {0.07} & {22.6} & {4.4} & {588.5} \\
\ch{{}^{19}F} & {3.4} & {70.1} & {73.6} & {0.05} & {21.9} & {5.3} & {721.4} \\
\ch{{}^{20}Ne} & {3.6} & {74.0} & {83.2} & {0.12} & {23.1} & {4.4} & {646.5} \\
\ch{{}^{21}Ne} & {3.8} & {77.5} & {83.1} & {0.07} & {21.8} & {5.6} & {756.1} \\
\ch{{}^{22}Ne} & {4.0} & {80.7} & {86.1} & {0.07} & {21.4} & {6.3} & {833.2} \\
\ch{{}^{22}Na} & {4.1} & {81.4} & {87.8} & {0.08} & {21.7} & {5.2} & {794.3} \\
\ch{{}^{23}Na} & {4.3} & {84.9} & {92.8} & {0.09} & {21.7} & {5.6} & {849.5} \\
\ch{{}^{24}Na} & {4.5} & {88.2} & {95.2} & {0.08} & {21.2} & {5.9} & {921.9} \\
\ch{{}^{24}Mg} & {4.5} & {88.8} & {102.0} & {0.15} & {22.6} & {5.1} & {839.7} \\
\ch{{}^{25}Mg} & {4.7} & {92.4} & {103.7} & {0.12} & {21.8} & {5.7} & {934.3} \\
\ch{{}^{26}Mg} & {5.0} & {95.6} & {106.8} & {0.12} & {21.5} & {6.2} & {1012.1} \\
\ch{{}^{28}Mg} & {5.4} & {101.5} & {109.1} & {0.07} & {20.3} & {6.4} & {1176.4} \\
\ch{{}^{26}Al} & {5.0} & {96.6} & {107.2} & {0.10} & {21.7} & {5.3} & {973.0} \\
\ch{{}^{27}Al} & {5.2} & {99.8} & {113.6} & {0.14} & {21.7} & {5.7} & {1034.3} \\
\ch{{}^{28}Si} & {5.5} & {103.6} & {121.5} & {0.17} & {22.1} & {5.4} & {1044.8} \\
\ch{{}^{29}Si} & {5.7} & {107.2} & {122.4} & {0.14} & {21.4} & {5.7} & {1161.4} \\
\ch{{}^{30}Si} & {6.0} & {110.5} & {125.0} & {0.13} & {21.0} & {6.0} & {1242.3} \\
\ch{{}^{31}Si} & {6.2} & {113.6} & {125.2} & {0.1} & {20.3} & {5.9} & {1348.4} \\
\ch{{}^{32}Si} & {6.4} & {116.6} & {128.4} & {0.1} & {20.1} & {5.9} & {1401.8} \\
\ch{{}^{31}P} & {6.2} & {114.6} & {129.5} & {0.13} & {20.8} & {5.5} & {1291.6} \\
\ch{{}^{32}P} & {7.5} & {117.9} & {146.9} & {0.25} & {19.6} & {4.7} & {1677.3} \\
\ch{{}^{33}P} & {6.7} & {121.1} & {133.0} & {0.1} & {19.9} & {5.6} & {1473.6} \\
\ch{{}^{32}S} & {6.5} & {118.4} & {136.0} & {0.15} & {21.0} & {5.1} & {1315.1} \\
\ch{{}^{33}S} & {6.7} & {122.0} & {135.7} & {0.11} & {20.1} & {5.3} & {1456.8} \\
\ch{{}^{34}S} & {7.0} & {125.4} & {138.7} & {0.11} & {19.9} & {5.4} & {1533.4} \\
\ch{{}^{35}S} & {7.2} & {128.5} & {138.4} & {0.08} & {19.2} & {5.1} & {1642.3} \\
\ch{{}^{36}S} & {7.4} & {131.6} & {140.9} & {0.07} & {18.9} & {4.9} & {1703.6} \\
\ch{{}^{35}Cl} & {7.3} & {129.4} & {143.9} & {0.11} & {19.8} & {4.7} & {1567.0} \\
\ch{{}^{36}Cl} & {7.5} & {132.8} & {144.3} & {0.09} & {19.2} & {4.5} & {1675.9} \\
\ch{{}^{37}Cl} & {7.8} & {136.8} & {151.2} & {0.11} & {19.4} & {4.0} & {1697.4} \\
\ch{{}^{36}Ar} & {7.7} & {136.0} & {146.7} & {0.08} & {18.9} & {4.3} & {1752.3} \\
\ch{{}^{37}Ar} & {7.5} & {133.2} & {151.5} & {0.14} & {20.1} & {4.1} & {1568.2} \\
\ch{{}^{38}Ar} & {8.0} & {140.2} & {154.6} & {0.1} & {19.2} & {3.9} & {1769.9} \\
\ch{{}^{39}K} & {8.3} & {144.2} & {160.9} & {0.12} & {19.3} & {2.7} & {1781.0} \\
\ch{{}^{40}Ca} & {8.6} & {148.0} & {168.8} & {0.14} & {19.6} & {1.5} & {1778.4} \\

\hline
\end{tabular}
\caption{CI-SM results characterising $B(E1)$ distributions of $sd$-shell nuclei close to the stability line computed in this work: Total strength ($S_0$), energy-weighted sum rule ($S_1$),
enhancement coefficient ($K$), centroid ($\bar S$) and width ($\Delta S$) of the distribution, (-2) moment of the photoabsorption cross section ($\sigma_{-2}$). Classical TRK sum rules ($S_1^{TRK})$ are added for comparison.}
\label{tab-results}
\end{table*}

Table \ref{tab-results} summarizes our theoretical results for all $sd$-shell nuclei near stability ($\tau > 1$ hour). Total strength $S_0$, the first moment $S_1$ and the TRK sum rule values are given for each of calculated nuclei. Those are followed by the extracted enhancement factor $K$, centroid $\bar S$ and width $\Delta S$ of the distributions.   
Additionally, we report in Table \ref{tab-results} the computed values of $\sigma_{-2}$. When using the non-regularized $E1$ operator,  
the $\sigma_{-2}$ values from our work coincide within a few percent with those from \cite{Orce2023}, despite different effective interactions employed and small differences in energy ranges used for evaluating $\sigma_{-2}$. This consistency highlights the robustness of the CI-SM description of $E1$ strengths. The results reported in the Table computed with an effective $E1$ operator derived here are obviously lower than those of Ref. \cite{Orce2023} using the bare $E1$ operator and would thus fit better $\sigma_{-2}$ values derived from experimental data, as can be anticipated from our results in Fig. \ref{PSF_SM_EXP_QRPA}.


\section{Electric dipole response in Neon chain\label{Neon}}
In this section, we present the results of our calculations along the Ne isotopic chain, with a detailed analysis of the dipole response in $^{26}$Ne. This nucleus has been extensively studied both theoretically \cite{Sieja-PRL2, Kimura2017, Martini-neon, Cao2005} and experimentally \cite{Gibelin2008}. Experimental work suggested the presence of the PDR mode, with $B(E1)$ strength below 10 ~MeV, accounting for approximately $4\%$ of the TRK sum rule.

The starting point of our discussion is defining what is meant by the "pygmy" dipole resonance: nowadays this term is frequently used for the concentration of the low-lying $E1$ strength, without implying any particular structure. The early interpretation of this mode in the three-fluid hydrodynamical model \cite{Mohan71} lead to two independent electric dipole resonances, one originating from the oscillation of all protons against all neutrons (GDR) and an energetically lower-lying mode where only the excess neutrons oscillate against a proton–neutron saturated core. The former mode was estimated to be more than two orders of magnitude stronger than the latter one (in $^{208}$Pb) which is in agreement with our present-day experimental knowledge \cite{Savran2013}.
The magnitude of the resonance being one indication, we shall further distinguish whether a low-energy peak can be classified as pygmy resonance or not. We assume here that the PDR is formed by low-energy peaks of the same structure, which should be clearly different from the structure of the GDR. With this definition we do not impose any particular character or collectivity on the PDR: note that different theoretical approaches predict systematically low-energy $E1$ strength 
but its collectivity and resonant nature are still debated and some authors prefer to use the term Pygmy Dipole Strength (PDS) rather than PDR \cite{Paar2005, Vretenar2012}. We will conclude from our calculations to which extent the PDR, as defined above, can be considered collective and associated with a neutron-skin oscillation. 

\begin{figure}
\begin{center}
\includegraphics[width=0.5\textwidth]{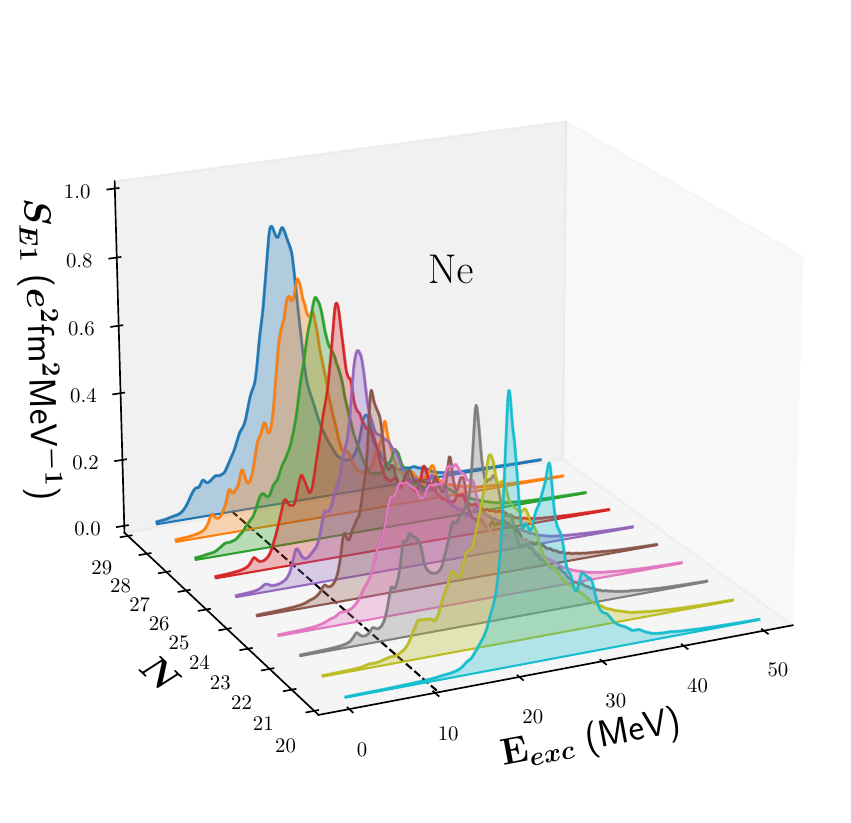}
\end{center}
\caption{Electric dipole strength functions obtained in the $1\hbar\omega$ shell-model calculations (Eq. \ref{Eq-SE1}) for even-even and even-odd neon isotopes with $N<20$.  
\label{fig-neon-ee}}
\end{figure}

In Fig. \ref{fig-neon-ee} the $E1$ response of even-even and even-odd Ne isotopes is presented, from the 
$N=Z$ line to $^{29}$Ne ($N=19$). We do not continue calculations towards larger $N$ as in the island of inversion region, i.e. at $N=20$,
the intruder configurations are supposed to appear in the ground states. Our approach in which the ground states are described as $0\hbar\omega$ excitations would be thus no longer valid. Starting with the lightest isotopes, one can observe the splitting of the GDR in $^{20}$Ne and $^{22}$Ne, compatible with a deformed ground state. The quadrupole moments resulting from the Hartree-Fock-Bogoliubov (HFB) calculations in the $sd$-shell with the USDb interaction and using the HF-SHELL solver \cite{HF-SHELL} are given in Tab. \ref{tab-qm}. Note that the intrinsic quadrupole moment evaluated in this manner only reflects the deformation of valence particles within the model space. Therefore, its value cannot be directly compared to experimental results or QRPA calculations. However, it serves as a useful reference for tracking the development of deformation along the Ne isotopic chain. This method consistently indicates the largest quadrupole collectivity near the $N=Z$ line, where the splitting of the GDR is observed, with a tendency towards sphericity as the neutron number increases. Precisely, the obtained deformation is the largest in $^{20-22}$Ne nuclei, predicted to have $\beta=0.5$ in the previous QRPA study with the Gogny forces \cite{Martini-neon}. 

\begin{table}[]
\caption{\label{tab-qm} Quadrupole moments resulting HFB calculations in the $sd$-shell valence space for Ne isotopes considered in this work.}
\begin{tabular}{cccc}
\hline
Nucleus & $Q_{20}$(e$^2$fm$^2$) & Nucleus & $Q_{20}$(e$^2$fm$^2$) \\
\hline
$^{20}$Ne & 15.2& $^{21}$Ne & 15.8 \\
$^{22}$Ne & 16.5& $^{23}$Ne & 14.1 \\
$^{24}$Ne & 11.2& $^{25}$Ne & 10.1 \\
$^{26}$Ne & 8.3 & $^{27}$Ne & 7.6 \\
$^{28}$Ne & 7.1 & $^{29}$Ne & 0.0 \\
\hline
\end{tabular}
\end{table}

In $^{20}$Ne we do not predict any $E1$ strength below 10 ~MeV. Interestingly, two low-energy $E1$ transitions were observed experimentally \cite{NNDC} at 5.78~MeV and 8.71~MeV with sizable $B(E1)$ strengths, indicating a possibility of isospin mixing in those low-energy states. While the $\Delta T=0$ $E1$ transitions are forbidden in $N=Z$ nuclei and cannot be reproduced in the present theoretical approach, we note that our model predicts the first 1$^-$ state at 5.77 ~MeV as well as the 6th $1^-$ state at 8.64 ~MeV, in very good agreement with energies of the experimentally observed states. The first $E1$ transition is predicted in CI-SM at 11.45 ~MeV, fitting very well the reported experimental transition at 11.27 ~MeV \cite{NNDC}. Passing the $N=Z$ line $E1$ strength appears below $10$~MeV and shifts towards lower and lower energy with increasing neutron number. 
The non-zero $B(E1)$ transitions in $^{22}$Ne and in $^{24}$Ne appear around 6.5  ~MeV but the total strength below 10~MeV 
remains small in both nuclei ($\sum B(E1)<0.006$e$^2$fm$^2$).
In $^{26}$Ne and $^{28}$Ne, after the first peak around 4.5~MeV, a bunch of strong transitions arises additionally in the energy range 7-9~MeV which may correspond to the PDR strength,
with the integrated $B(E1)$ of 0.28e$^2$fm$^2$ in $^{26}$Ne and 0.33e$^2$fm$^2$ in $^{28}$Ne. One may note that a better agreement with the experimental value in $^{26}$Ne (0.49$\pm$0.16e$^2$fm$^2$) is obtained without applying the newly estimated effective charge (0.44e$^2$fm$^2$). However, the use of the effective charges chosen in this work provides only an overall scaling of the $B(E1)$ values and, as such, does not play any role in the interpretation of different modes. 

The centroids of the PDR are located at 8.6 ~MeV in $^{26}$Ne and 7.95 ~MeV in $^{28}$Ne, which are in close alignment with the predictions from the quasiparticle relativistic random phase approximation (QRRPA) \cite{Cao2005}, where the dipole strength below 10 ~MeV is centered at 8.3 ~MeV for $^{26}$Ne and 7.9 ~MeV for $^{28}$Ne. It is worth noting that the QRPA framework with Gogny forces \cite{Martini-neon} also predicts low-energy strength in both nuclei, albeit at slightly higher energies, with the first peak occurring above 10 ~MeV. Furthermore, a PDR centered at 8.5 ~MeV in $^{26}$Ne was identified using shifted-basis antisymmetrized molecular dynamics (AMD) with generator coordinate method (GCM) \cite{Kimura2017}, again showing a close agreement in excitation energy with the CI-SM results. In the following subsection, we will examine in detail the properties of the low-energy strength in $^{26}$Ne and refer to these theoretical approaches whenever possible.

Finally, the odd-even Ne isotopes are shown in the Figure for completeness. The general features of the electric dipole response are very similar to that of even-even systems, with sizable $E1$ strength accumulating at low excitation energies in neutron-rich isotopes with integrated strength below $10$ MeV of $0.18$, $0.34$ and $0.34$ $e^2$fm$^2$ for $^{25}$Ne, $^{27}$Ne and $^{29}$Ne, respectively. 

\subsection{Electric dipole response of the ground-state in $^{26}$Ne\label{neon26}}
\begin{figure}
\includegraphics[width=0.5\textwidth]{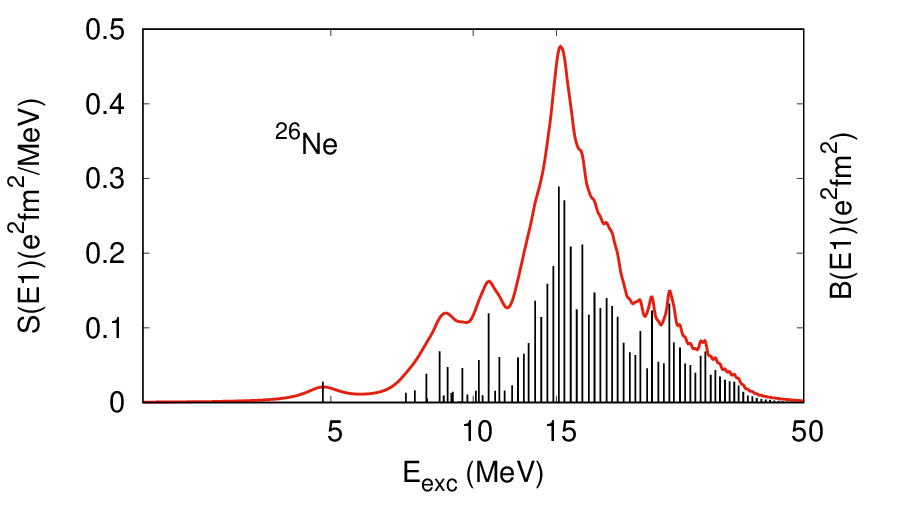}
\caption{Electric dipole response of $^{26}$Ne computed in this work. Discrete distribution of $B(E1)$ values (black peaks) obtained via Lanczos strength function method with 300 iterations is shown as well as the folded microscopic strength $S_{E1}$ (red curve). \label{fig-26ne}}
\end{figure}

We will focus now our discussion on the $E1$ strength taking as example $^{26}$Ne. The dipole response of the ground state of this nucleus is detailed in Fig. \ref{fig-26ne}, where discrete $B(E1)$ values 
obtained with the Lanczos strength function method as well as the microscopic strength after the folding procedure, Eq. \ref{Eq-SE1}, are plotted. One can distinguish a first peak around 5~~MeV followed by two structures centered at 8.5~MeV and 11.5~MeV, and finally the GDR, 
with large peaks appearing above 15~MeV and the centroid of the distribution (10-50~MeV) at 19.4~MeV. 
The GDR appears thus lower in the CI-SM than in QRRPA \cite{Cao2005}
(22.32~MeV) and GCM \cite{Kimura2017} (22.7-24.5~MeV) approaches. 

\begin{figure*}[hbt]    
    \includegraphics[width=0.24\textwidth]{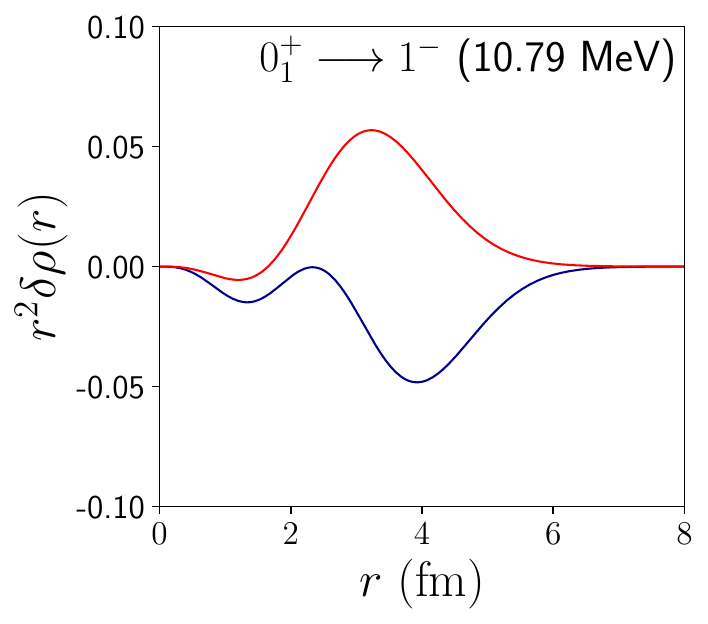}
    \includegraphics[width=0.24\textwidth]{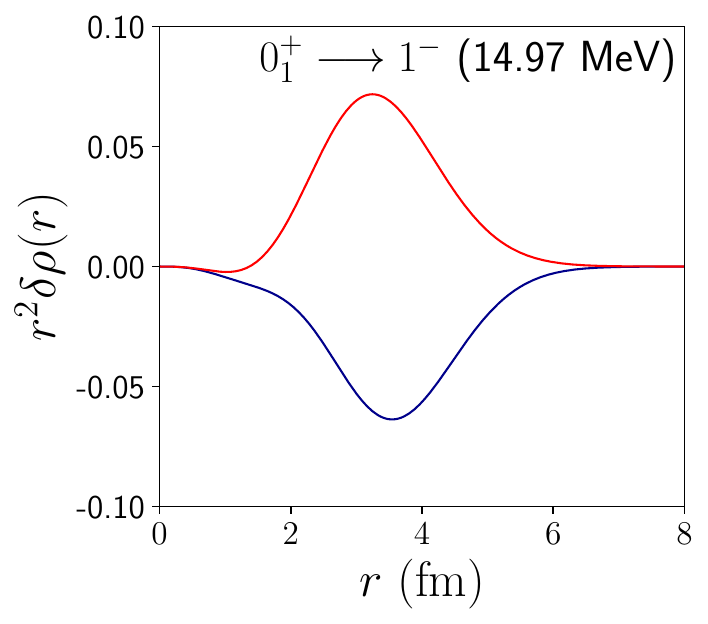}
    \includegraphics[width=0.24\textwidth]{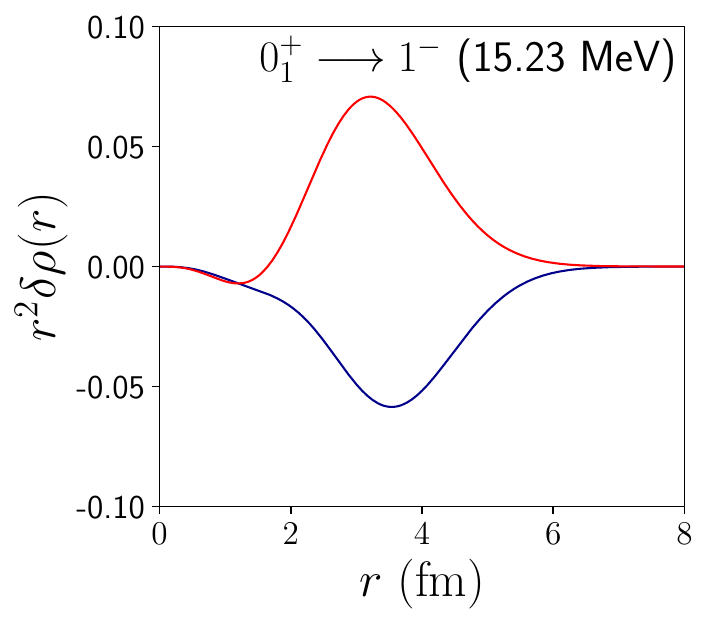}
    \includegraphics[width=0.24\textwidth]{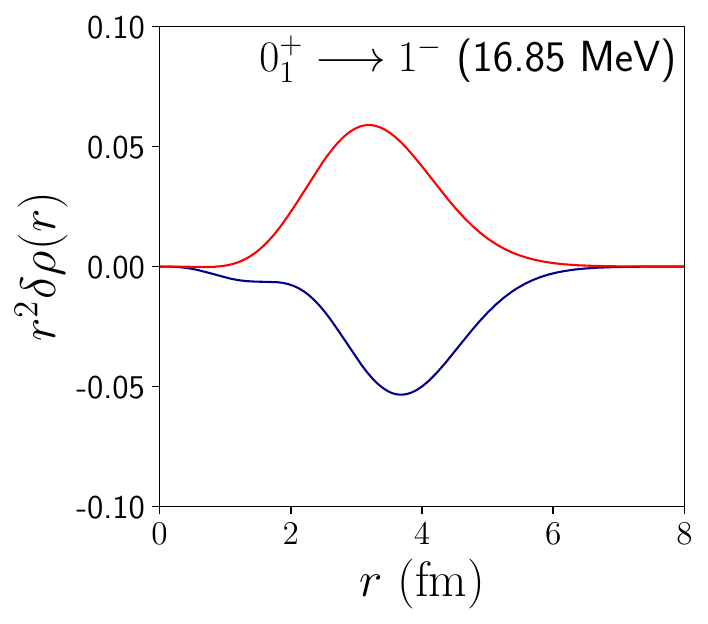}
    \\
\hfill
    \includegraphics[width=0.24\textwidth]{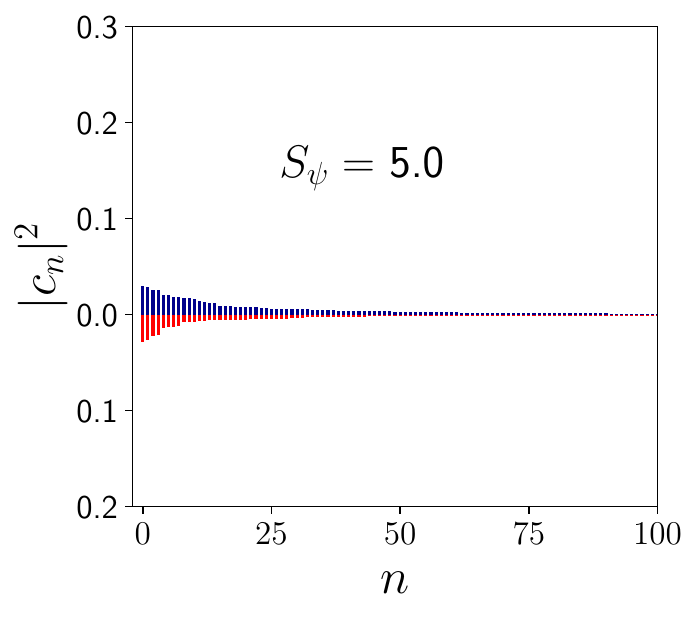}
    \includegraphics[width=0.24\textwidth]{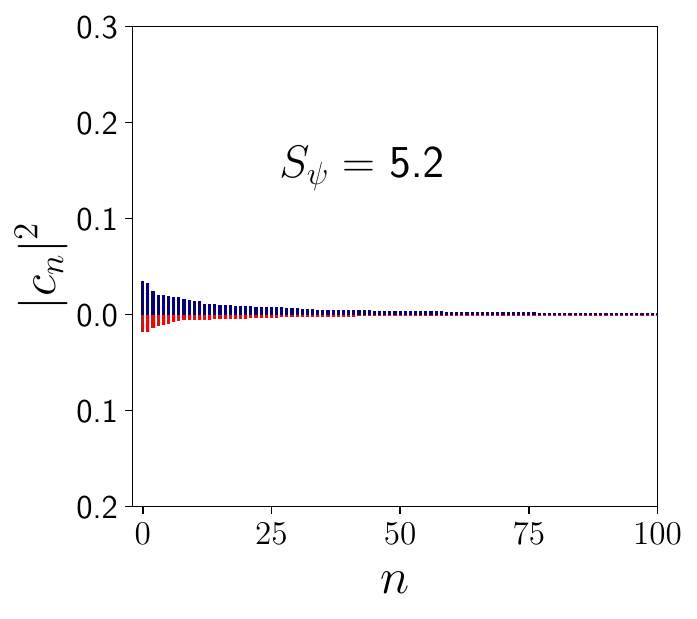}
    \includegraphics[width=0.24\textwidth]{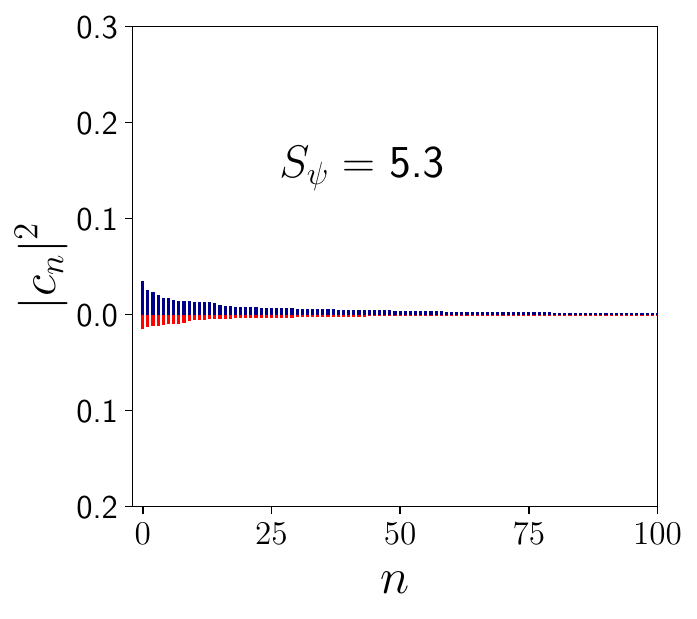}
    \includegraphics[width=0.24\textwidth]{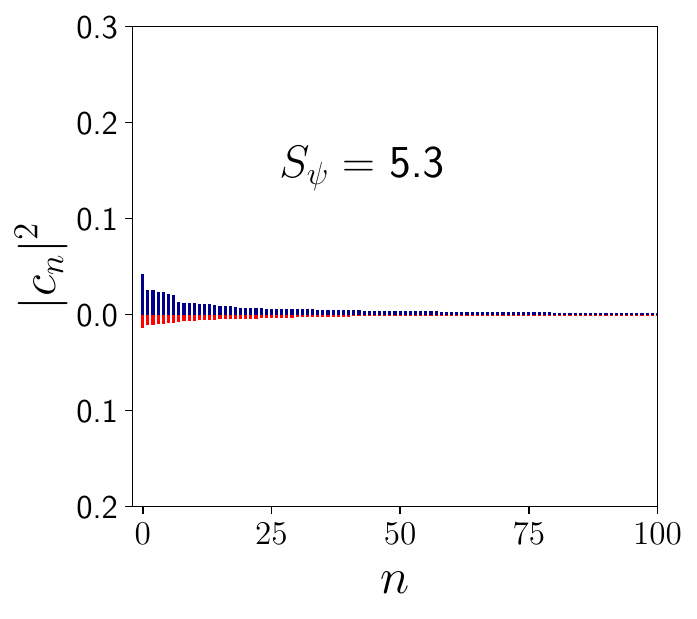}
\hfill
\caption{Neutron (blue) and proton (red) transition densities of the selected states in the GDR region (top panels) and the corresponding wave-functions decomposition into different configurations (bottom panels). The excitation energy (in paranthesis) and the configuration entropy are given for each considered $1^-$ states.\label{fig-GDR}}
\end{figure*}
With the chosen effective charges values our calculations always exhaust the TRK sum rule, see Tab. \ref{tab-results}, thus we consider the GDR is fully collective and may serve us as a reference to analyze the collective properties of the low-energy states. In Fig. \ref{fig-GDR} shown are the neutron and proton transition densities computed for selected states in the GDR region and in its tail. The spreading of the corresponding wave-functions among various configurations is shown in the same Figure.  

As expected, the transition densities exhibit 0-node, purely isovector character with identical behavior in the GDR states, while a slight variation of the transition density is observed in the tail of the GDR, still maintaining the isovector nature. The wave functions of the GDR states are widely spread over many proton-neutron configurations, there are no components with probabilities exceeding 5$\%$. The value of $S_\Psi$ of 5.3 in the main GDR peaks decreases slightly in the tail, still being larger than for the low-energy $1^-$ states, as will be shown below. The bottom panel of Fig. \ref{fig-cumul} illustrates the contributions to the 
$B(E1)$ value from the $X_{k_{\alpha}k_{\beta}}$ proton and neutron components in Eq. \ref{eq-cumul}:
predictably, the GDR strength originates from a coherent summation of proton and neutron components that are all in phase.

\begin{figure}
\begin{center}
 \includegraphics[width=0.3\textwidth]{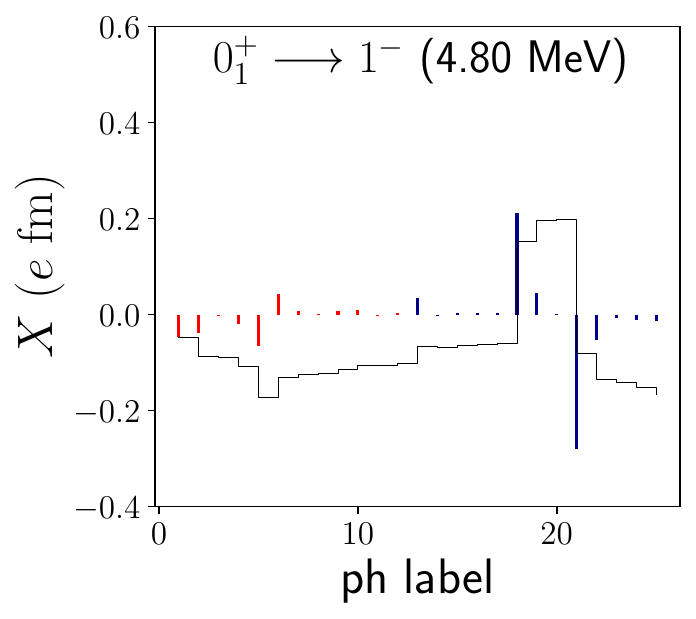}
    \includegraphics[width=0.3\textwidth]{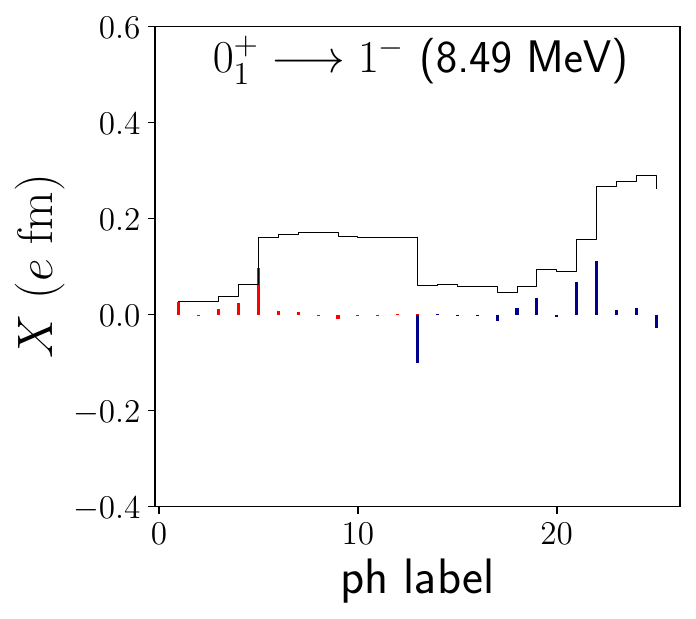}
   \includegraphics[width=0.3\textwidth]{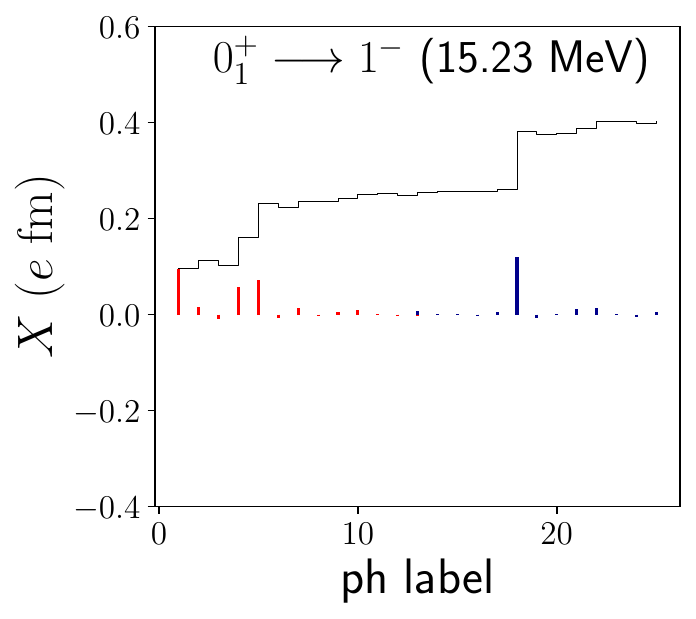}  
   \end{center}
\caption{Neutron (blue) and proton (red) particle-hole contributions $X$ from Eq. \ref{eq-cumul} and their cumulative sum (black line) for the first excited $1^-$ state (upper panel) and selected PDR (middle panel) and GDR peaks (bottom panel).\label{fig-cumul}}
\end{figure}

Having established the isovector and collective character of the GDR and of its tail, we shall now examine the $1^-$ states located below 10~MeV.
The CI-SM predicts in total 14 states with non-zero $B(E1)$ transitions, 2 of them in the first peak centered at 5~MeV, 12 within the PDR region, as listed in Table \ref{tab-PDR}. Clearly, none of the previous theoretical calculations in $^{26}$Ne yields such a large fragmentation of the low-energy strength, though the predicted integrated $B(E1)$ below 10~MeV value is similar in all calculations. It is known that going beyond the QRPA is necessary to increase the fragmentation, which is not the case of Refs. \cite{Martini-neon, Cao2011}. The GCM from Ref.~\cite{Kimura2017} could be more precise for the low-energy levels and indeed seems to be more consistent with the CI-SM results, as it also predicts two low-lying states around 5~MeV. Still, there are only 4 $1^-$ states in the PDR region against 12 in CI-SM. It can be anticipated simply from the number of the obtained excited states that the CI-SM predicts more complex structures than the remaining approaches.

\begin{table}
\caption{Low-energy $1^-$ states computed in the present CI-SM approach with their corresponding $B(E1)$ strengths from the ground state ($0_1^+$) and first excited $0^+$ state ($0_2^+$).\label{tab-PDR}}
\begin{tabular}{cccc} 
\hline
  & {E$_{\textrm{exc}}$} & {$B(E1;0^+_1\rightarrow1^-)$} & {$B(E1;0^+_2\rightarrow1^-)$}\\
  & (~MeV) & (10$^{-2}$e$^2$fm$^2$) & (10$^{-2}$e$^2$fm$^2$)\\
\hline
     $1^-_1$ & 4.80&  2.78  &  0.94\\
     $1^-_2$ & 5.54&  0.0025 &  0.17\\
     $1^-_3$ & 7.20&  1.28 & 0.78\\  
     $1^-_4$ & 7.52&  1.63 & 0.49\\    
     $1^-_5$ & 7.95&  3.84 & 0.002\\      
     $1^-_6$ & 7.99&  0.54 & 0.21\\ 
     $1^-_7$ & 8.49&  6.84 & 0.56 \\         
     $1^-_8$ & 8.66&  0.90  & 0.04\\     
     $1^-_9$ & 8.82&    4.73 & 0.07 \\       
     $1^-_{10}$& 8.99&  1.27 & 0.18\\      
     $1^-_{11}$& 9.05&  1.41 & 0.01\\      
     $1^-_{12}$& 9.35&  0.14 & 0.10\\      
     $1^-_{13}$& 9.48&  4.62 & 0.24\\      
     $1^-_{14}$& 9.72 &  1.06& 0.49\\   
\hline
\end{tabular}
\end{table}

\begin{figure*}[hbt]
\includegraphics[width=0.19\textwidth]{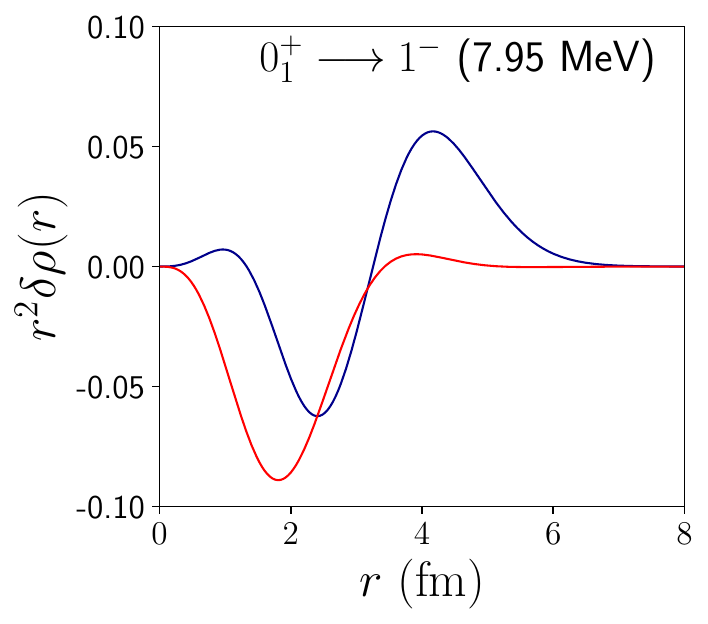}
\includegraphics[width=0.19\textwidth]{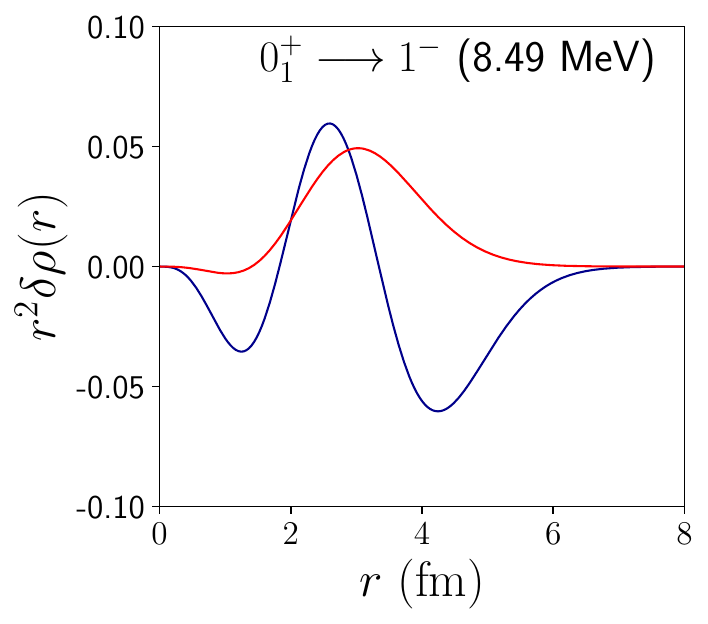}
\includegraphics[width=0.19\textwidth]{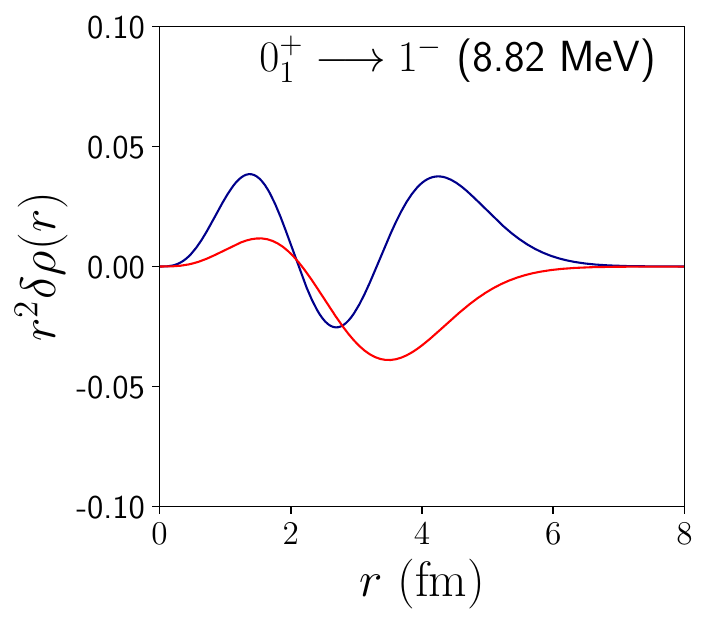}
\includegraphics[width=0.19\textwidth]{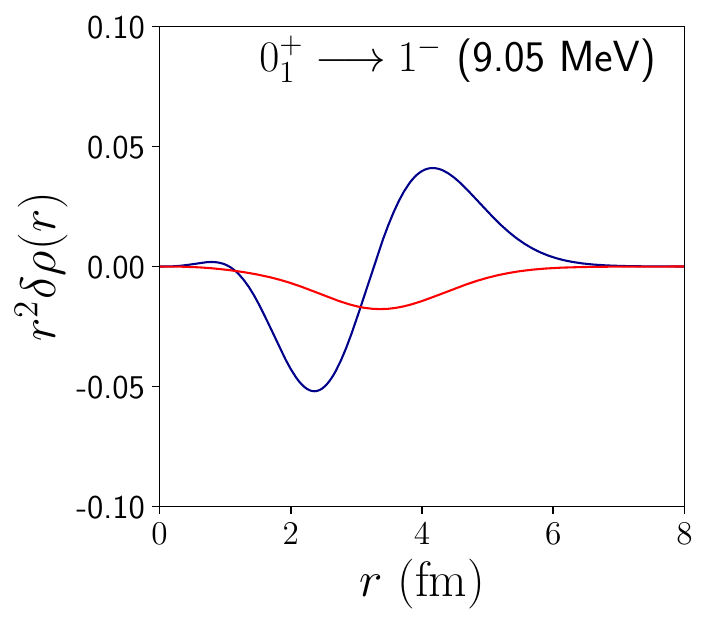}
\includegraphics[width=0.19\textwidth]{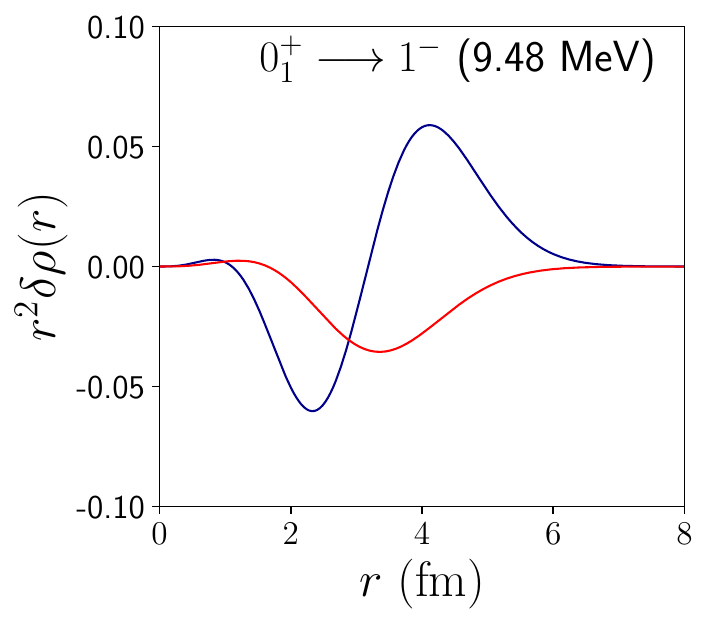}
\\
\hfill
\includegraphics[width=0.19\textwidth]{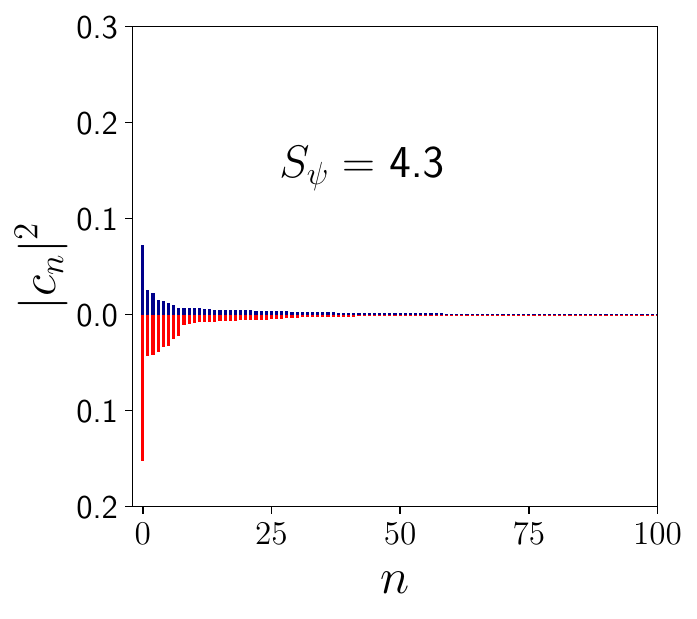}
\includegraphics[width=0.19\textwidth]{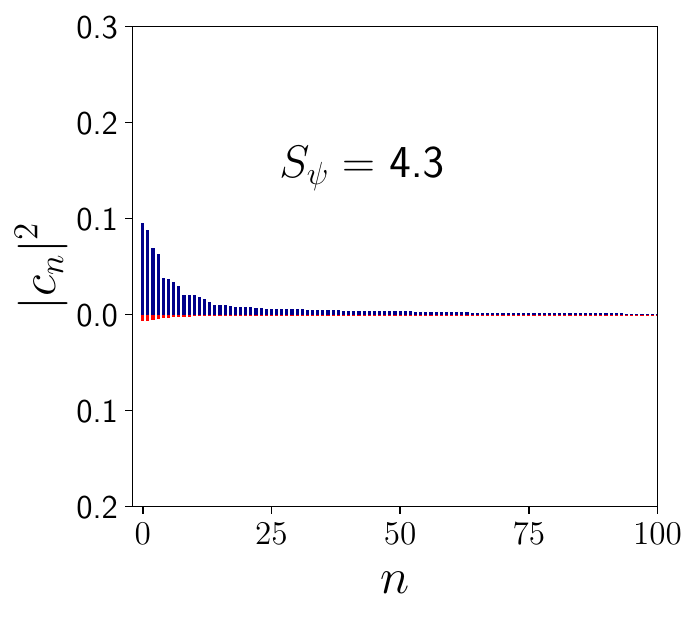}
\includegraphics[width=0.19\textwidth]{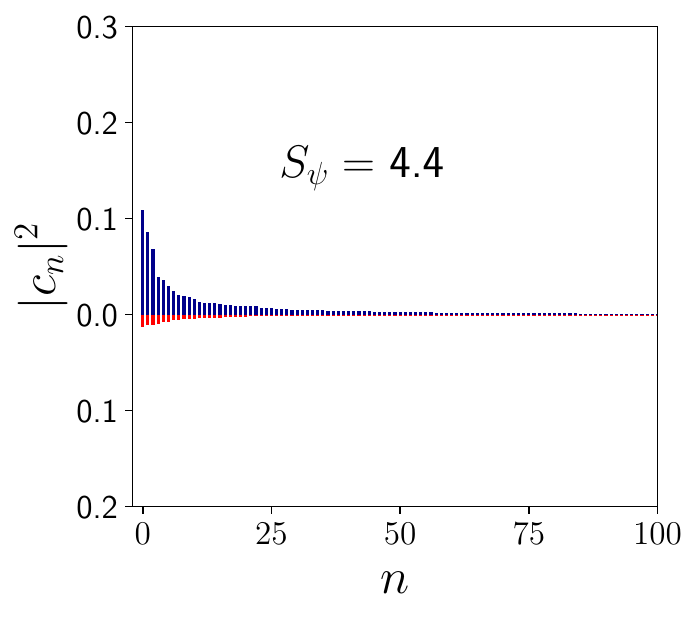}
\includegraphics[width=0.19\textwidth]{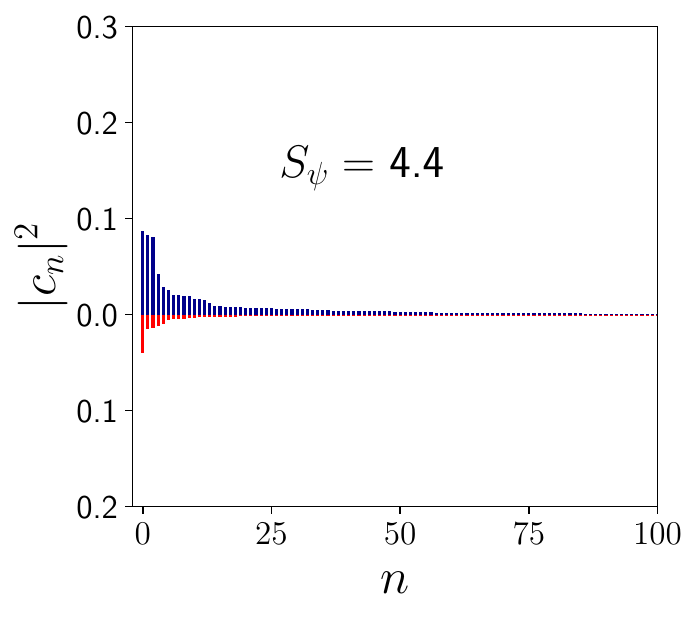}
\includegraphics[width=0.19\textwidth]{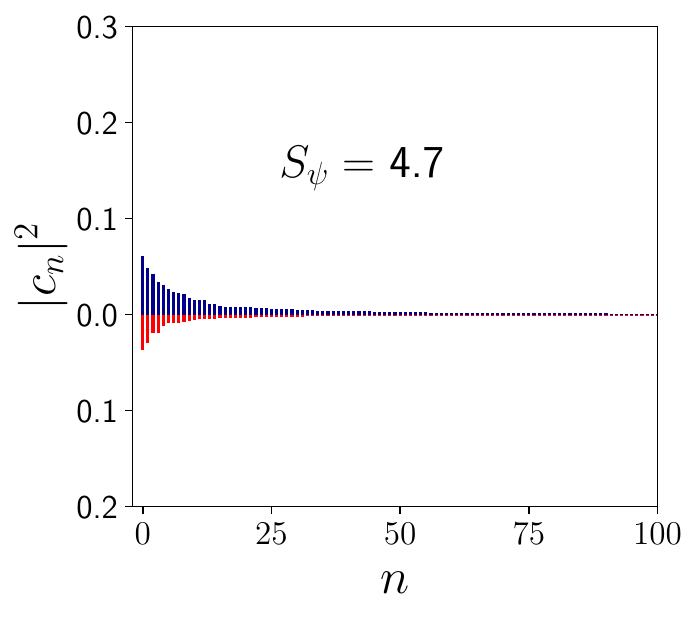}
\hfill
\caption{The same as in Fig. \ref{fig-GDR} but for $1^-$ states of $^{26}$Ne in the PDR region. \label{fig-PDR}}
\end{figure*}

This fact is illustrated in Fig. \ref{fig-PDR} where, as in the case of the GDR, we depict the neutron and proton transition densities of 5 $1^-$ states with the strongest $B(E1)$ values and their corresponding wave functions. As in the case of the GDR, the transition densities of the states at the extremities of the PDR (not shown here) exhibit slightly different behavior, while those close to the PDR centroid are fairly similar and share the common feature of a large neutron oscillation at the edge of the nucleus. The wave functions are still widely spread with several dominant components reaching up to 10$\%$ and $S_\Psi=4.3-4.7$, thus a reduction of entropy of $11-20\%$ with respect to the GDR. Obviously, the collectivity of the PDR states has to be lower to result in a lower $B(E1)$ value: as illustrated in the middle panel of Fig. \ref{fig-cumul},
the $B(E1)$ value arises as a sum of several neutron components with two contributions that are not in phase. Such 1-2 incoherent contributions are observed in all $1^-$ states between 7 and 10~MeV. As can be expected for a soft mode, only a group of nucleons participates in the collective motion, leading to a more complex pattern of the p-h excitations than for the GDR.

Except for the 7.95~MeV state which has a large proton contribution, other states between 7 and 10~MeV are all dominated by neutron excitations. It follows from the above discussion and from Fig. \ref{fig-PDR} that the CI-SM supports the "classical" PDR picture, i.e. of the resonant excitation (in the 8-9~MeV energy region) which can be associated with the neutron-skin oscillation. 
   
\begin{figure}
\includegraphics[width=0.23\textwidth]{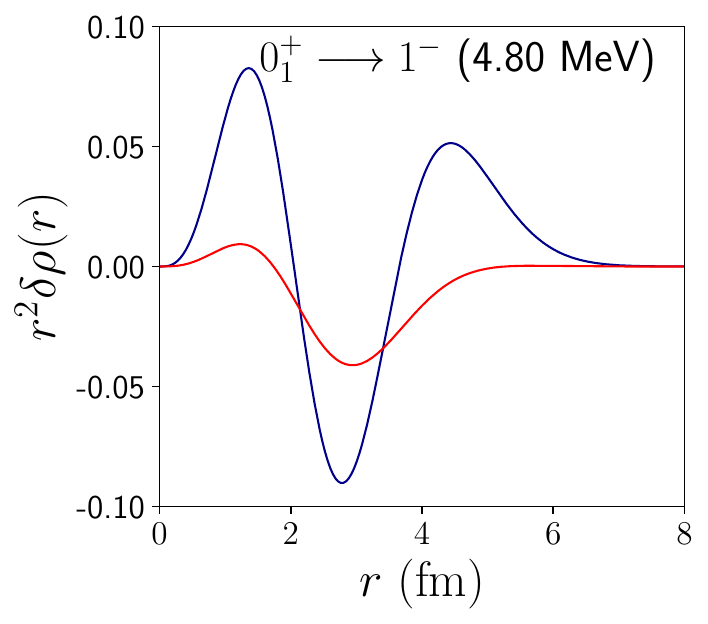}
\includegraphics[width=0.23\textwidth]{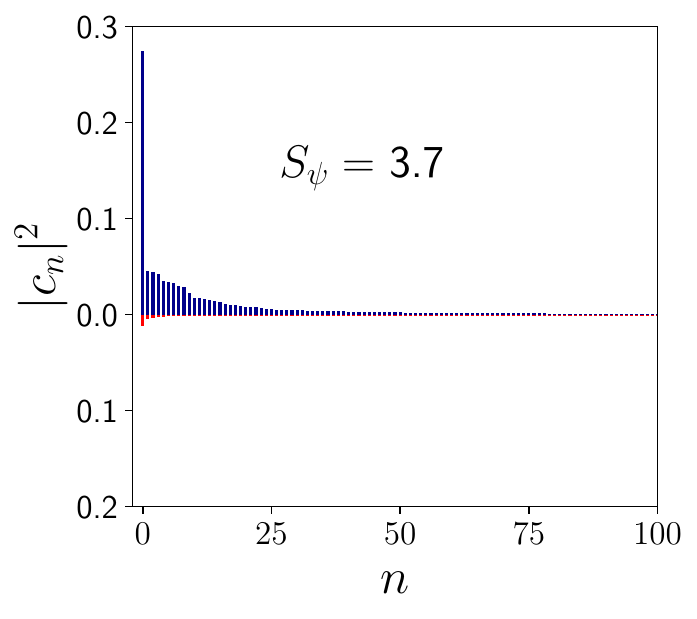}

\caption{The same as in Figs. \ref{fig-GDR} and \ref{fig-PDR} but for the first-excited 1$^-$ state in $^{26}$Ne. \label{fig-1st}}
\end{figure}

To complete the analysis of the low-energy strength, we pay attention to the first peak at 4.8~MeV which has a sizeable $B(E1)$ value. Its transition density and wave function are plotted 
in Fig. \ref{fig-1st} while the particle-hole contributions to the $B(E1)$ values are shown in the top panel of Fig. \ref{fig-cumul}. All the quantities are different from those
in the PDR region; neither the 4.8~MeV state hold the same dipole collective character as the PDR, with one contribution standing above the others in the wave function ($\nu s_{1/2}^{-1}p_{3/2}^1$), the lowest $S_\psi$ value among the studied states and large cancellations between proton and neutron p-h components. A closer look to the structure of the wave function reveals it is dominated by protons coupled to $0^+$ (48$\%$) or $2^+$ (40$\%$) coupled to neutron's $1^-$ (63$\%$) or $3^-$ (23$\%$).
In the PDR region $1^-$ states arise from the combination of many more components, where both neutrons and protons couple to positive or negative parity states. Substantial contribution of the $\nu 3^-\otimes \pi 2^+$ in the first $1^-$ state results in its large $B(E2; 3^-\rightarrow 1_1^-)$ value of $\sim$10~W.u., which decreases to 1 W.u. for the $1^-_2$ and to 0.1~W.u. for $1^-$ states above 7~MeV. 

The difference of structure of those first excited states is further reflected by the computed spectroscopic factors for one neutron removal from $^{26}$Ne excited states, presented in Tab. \ref{sf}. The first $1^-$ state has clearly the largest overlap with the $^{25}$Ne ground state $1/2^+$. The second $1^-$, with the lowest $B(E1)$ value, has the largest overlap with the $5/2^+$ state. A change of structure is seen starting with the 3rd excited state, marking the beginning of the PDR region. The spectroscopic factor values drop substantially, as may be expected from the complexity of wave-functions established for the PDR states. The values are of the same order for the excited $3/2^+$ and $5/2^+$ states as for the ground state of $^{25}$Ne. This is consistent with the observation that the PDR does not necessarily decay to the ground state, but to the excited state of $^{25}$Ne. We note those results are also in qualitative agreement with the GCM calculations from Ref. \cite{Kimura2017}. 

\begin{table}
\caption{Spectroscopic factors in $^{25}$Ne$(J^\pi)\otimes nlj$ channels for selected low-lying $1^-$ states in $^{26}$Ne.}
\label{sf}
\begin{center}
\begin{tabular}{SSSSS}
\hline
\addlinespace[1ex]
{} & {$1/2^+_1 \otimes 1p_{3/2}$} & {$1/2^+_1 \otimes 1p_{1/2}$} & {$3/2^+_1 \otimes 1p_{3/2}$} & {$3/2^+_1 \otimes 0f_{5/2}$} \\
\addlinespace[1ex]
\hline
\addlinespace[1ex]
{$1_1^-$} & {0.45} & {0.02} & {0.06} & {0.01} \\
{$1_2^-$}  & {0.0} & {0.18} & {0.16} & {0.0}\\
{$1_3^-$} & {0.1} & {0.01} & {0.0} & {0.0}\\
{$1_4^-$} & {0.0} &{0.01} & {0.08} & {0.01}\\
{$1_5^-$} & {0.04} & {0.0} & {0.04} & {0.0} \\
{$1_7^-$} & {0.02} & {0.13} & {0.01} & {0.0} \\
{$1_9^-$} & {0.04} & {0.07} & {0.12} & {0.0} \\
{$1_{11}^-$} & {0.04} & {0.10} & {0.08} & {0.0} \\
\addlinespace[1ex]
\hline
{} & {$3/2^+_1 \otimes 1p_{1/2}$} & {$5/2^+_1 \otimes 0f_{7/2}$} & {$5/2^+_1 \otimes 1p_{3/2}$} & {$5/2^+_1 \otimes 0f_{5/2}$} \\
\addlinespace[1ex]
\hline
\addlinespace[1ex]
{$1_1^-$} & {0.03} & {0.15} & {0.05} & {0.0} \\
{$1_2^-$} & {0.0} &  {0.01} & {0.32} & {0.01}\\
{$1_3^-$} & {0.01} & {0.03} & {0.0} & {0.01} \\
{$1_4^-$} & {0.11}& {0.0} & {0.0} & {0.0} \\
{$1_5^-$} & {0.0}  & {0.0} & {0.02} & {0.0} \\
{$1_7^-$} & {0.13} & {0.01} & {0.06} & {0.0} \\
{$1_9^-$} & {0.07} & {0.0} & {0.05} & {0.0} \\
{$1_{11}^-$} & {0.07} & {0.0} & {0.05} & {0.0} \\
\addlinespace[1ex]
\hline
\end{tabular}
\end{center}
\end{table}

\subsection{Electric dipole response of the excited $0^+$ state in $^{26}$Ne\label{neon26}}
In a previous work \cite{Sieja-PRL2} we have addressed the question of the validity of the Brink-Axel hypothesis in the PDR region based on CI-SM calculations within the present framework. Note that the Lanczos strength function method permits getting the strength 
distribution of any operator on any initial state, which offers 
the opportunity to test the Brink hypothesis, stating that the energy-smoothed photoabsorption cross section should not depend on the initial state. In Ref. \cite{Sieja-PRL2} the dipole photoresponse of Ne nuclei for ground and excited states of various spins was thus examined, proving that the EWSR computed in the 0-50~MeV range is fairly independent of the initial state, as it is dominated by the GDR. However, for the EWSR evaluated in the low-energy range, which is typically considered e.g. in the calculations of particle-capture cross sections in astrophysical settings, differences between the ground state and excited states were found, indicating there is no PDR built on excited states. This effect seems in odds with the majority of theoretical finite-temperature approaches, which typically predict more low-energy strength with increasing temperature due to thermal unblocking of extra states to excitation and broadening of the Fermi surface \cite{elena-Mo, Wibowo2018, Yuksel, Kaur2024}. 

In the CI-SM approach the redistribution of the dipole strength with excitation energy of initial states (up to 7~MeV) was detected in $^{26,27,28}$Ne nuclei. The dipole response computed on excited states exhibited a much smoother low-energy trend, without strong transitions to the $1^-$ states in the PDR region, leading to an overall lower EWSR in the $E_\gamma=0-10$~MeV range.  This fact is illustrated in Table \ref{tab-PDR}, where the $B(E1)$ strength is reported from the ground state $0^+_1$ and from the first excited $0^+_2$ (located at 4.3~MeV) to the first 14 $1^-$ states. One can note that all the transitions from $0^+_2$ to states between 7 and 10~MeV of excitation energy are smaller and some are severely quenched. Of course, this could be compensated by stronger transitions above 5~MeV in $\gamma$ energy, but apparently the strength computed on the excited $0^+$ state raises in a steady way. Altogether, the EWSR for $E_\gamma$=0-10~MeV amounts to 2.56e$^2$fm$^2$~MeV for the ground state and to 1.94e$^2$fm$^2$~MeV for the excited state, so that the reduction of $\sim24\%$ is observed. 

\begin{figure}
\includegraphics[width=0.25\textwidth]{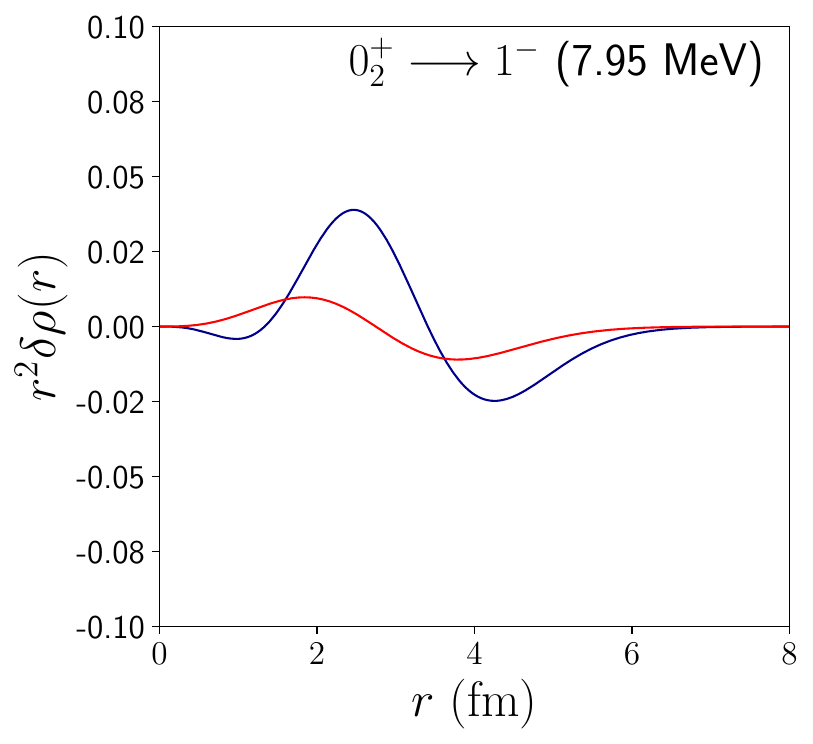}\includegraphics[width=0.25\textwidth]{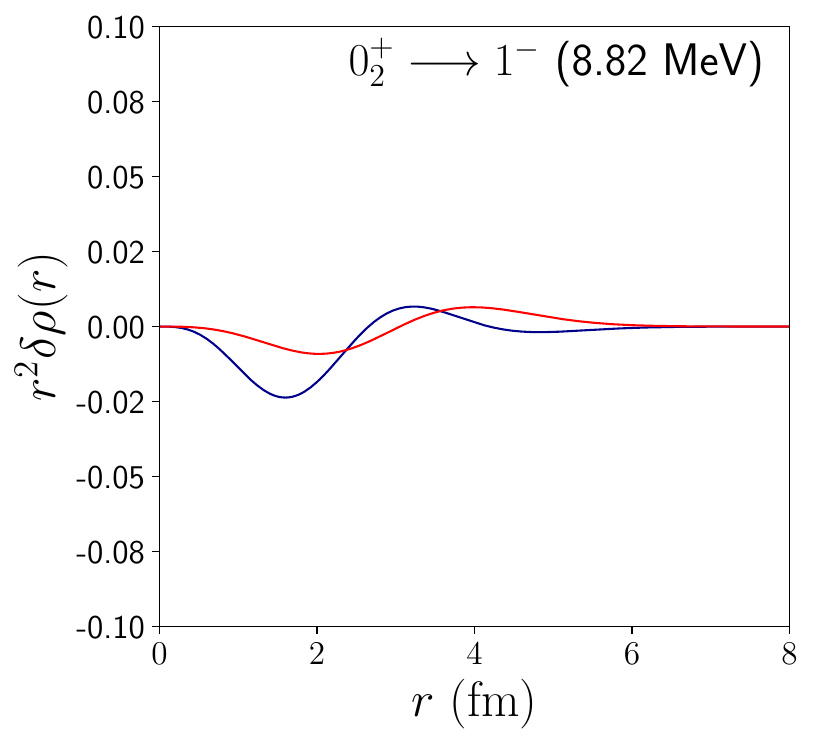}
\caption{Neutron (blue) and proton (red) transition densities connecting the first excited $0^+$ and 1$^-_5$ and 1$^-_9$ states in $^{26}$Ne. See text for more details. \label{fig-dens-exc}}
\end{figure}

The $0^+$ ground state in $^{26}$Ne is formed by two neutrons in the $s_{1/2}$ shell. The second excited $0^+_2$ state in $^{26}$Ne is composed of 2 neutrons in the $d_{3/2}$ orbital. One can expect different matrix elements will thus be dominating in the computed transitions to low-energy $1^-$ states. In Fig. \ref{fig-dens-exc} we display the computed transition densities for the cases from Tab. \ref{tab-PDR} that show the largest difference of $B(E1)$ value, namely $1_5^-$ at 7.95~MeV and $1^-_9$ at 8.82~MeV, to be compared to their respective transition densities to the ground state in Fig. \ref{fig-PDR}.   
The transition densities in Fig. \ref{fig-dens-exc} are much flatter than those in Fig. \ref{fig-PDR}. The first examined transition density (7.95~MeV state) contains large $\nu d_{3/2}\rightarrow\nu1p_{3/2}$ OBTD instead of the $\nu d_{5/2}\rightarrow\nu 1p_{3/2}$ which is substantial in the transitions to the ground state. Also, in the 8.82~MeV peak, the OBTD to the excited $0^+$ is dominated by the neutron $d_{3/2}\rightarrow p$ component, which was negligible in the OBTD to the ground state. In both states, there is no large neutron density at the edge of the nucleus that could correspond to the skin oscillation. Further difference with respect to the PDR on the ground state is observed in the proton excitation, which is strongly reduced in the transitions to the excited state.  
A question remains if the observed redistribution of the low-energy $E1$ strength depends on the particular structure of the $0^+$ states in the $sd$-shell and the proton-neutron interactions of the PSDPF Hamiltonian. As a quick check, we have performed calculations in the $sd-pf-gds$ CI-SM framework following Ref. \cite{Sieja-PRL}. The effective interaction employed predicts an enhanced low-energy $E1$ strength in the PDR region in $^{48}$Ca 
and the reduction of the EWSR in the 0-10 MeV range between $0^+$ g.s. and first excited $0^+$ state by a very similar amount as in $^{26}$Ne. Although the effective interaction from \cite{Sieja-PRL} is not as precise as the PSDPF 
for the low-energy states and no systematics of dipole response of $pf$-shell nuclei have been performed yet, we find it appealing that the preliminary results for the low-energy strength in the $f_{7/2}$-shell nuclei seem to follow the same tendencies, indicating the predicted PDR is softened when computed on excited states. 

\section{Conclusions \label{CONC}}
We have performed systematic CI-SM calculations of the electric dipole response of the $sd$-shell nuclei
which can be reasonably described in the $1\hbar\omega$ model space. Our results show good agreement with the experimental data for the position of the GDR peak, width of the resonance and its shape, which has a clear advantage over the available QRPA calculations, even with empirical adjustments. However, the model significantly overshoots the experimental data, a trend observed in previous shell-model studies as well. We attribute this discrepancy to the use of a non-regularized $E1$ operator. As is common in truncated model-space calculations, we propose a straightforward prescription for an effective charge in the dipole operator to improve agreement with the available photoabsorption data. Nonetheless, it would be valuable to revisit the choice of CI-SM Hamiltonians and effective charges as more experimental data becomes available, including low-energy $E1$ transitions. 
In addition to the systematic calculations of $E1$ response, we investigated the dipole strength in Ne isotopes, focusing on $^{26}$Ne. Our results are overall compatible with previous theoretical studies within shifted-basis AMD+GCM and QRPA approaches with Gogny forces and with QRRPA in their predictions of the low-energy dipole strength in $^{26,28}$Ne, though larger fragmentation of the $E1$ strength at low energies is obtained in CI-SM. In order to answer the question about the nature of this low-energy strength we analyzed the transition densities and wave functions of $1^-$ states in $^{26}$Ne. We conclude that states located between 8 and 9 ~MeV have distinct structure, different from the GDR tail and from the lowest-energy excitations. They share common features among them, indicating a resonant nature. The PDR states exhibit lower dipole collectivity than those in the GDR region, but are more collective than the lowest peaks appearing in the calculations. The computed transition densities support the picture of oscillating neutron skin in the PDR region and show the isovector behavior of the resonance at the edge of the nucleus. This is the first time such insight into the PDR structure is provided from the CI-SM perspective and we find it valuable to continue this type of studies. Future developments we will carry will extend the analysis of the PDR modes to Ca nuclei and to the role of the isoscalar dipole response. The systematic results on $E1$ PSF in $sd$-shell nuclei from this
work will be extended to the evaluation of the $M1$ mode.
Both $E1$ and $M1$ strengths will be used to estimate the photodisintegration cross sections 
with the TALYS code \cite{TALYS}, before being applied to astrophysical studies. The corresponding systematic cross section calculation and application to UHECR is currently under study.

\section{Acknowledgments}
This work of the Interdisciplinary Thematic Institute QMat, as part of the ITI 2021-2028 program of the University of Strasbourg, CNRS and Inserm, was supported by IdEx Unistra (ANR 10 IDEX 0002), and by SFRI STRAT’US project (ANR 20 SFRI 0012) and EUR QMAT ANR-17-EURE-0024 under the framework of the French Investments for the Future Program.

\bibliographystyle{apsrev}

\end{document}